\documentclass[preprint,12pt]{article}
\usepackage{graphicx}  
\usepackage{dcolumn}   
\usepackage{bm}        
\usepackage{amssymb}   
\usepackage{subfigure}
\usepackage{amsmath,amsfonts}
\usepackage{setspace}
\usepackage{authblk}
\usepackage[utf8]{inputenc}
 
\begin{document}

\title{Effect of cross-coupling on the phase behavior in a biaxial nematic: Insights
from Monte Carlo studies}
\author{B. Kamala Latha, K. P. N. Murthy, and V. S. S. Sastry}

\affil{School of Physics, University of Hyderabad, Hyderabad 500046,
            Telangana, India}

\begin{abstract}
 Phase sequences of the biaxial nematic liquid crystal in the interior of 
 the essential triangle are studied with Wang Landau sampling. The evidence
 points to the existence of an intermediate unixial phase with low biaxiality in the 
 isotropic to biaxial nematic phase sequence. 
\end{abstract} 
 \maketitle{}
 
\section{Introduction}
\begin{figure}[]
\centering{
\subfigure[]{\includegraphics[width=0.45\textwidth]{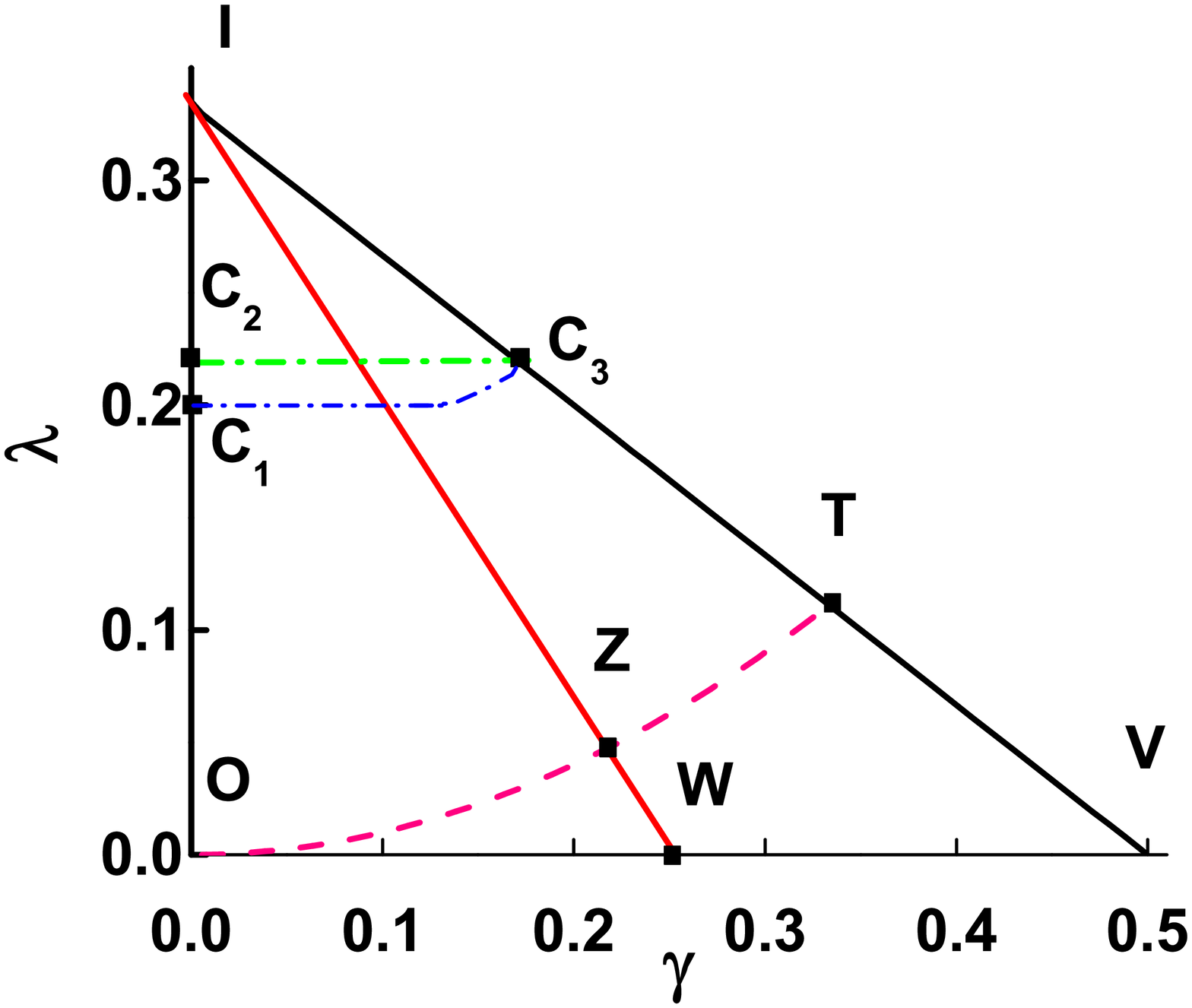}
\label{fig:1a}}
\subfigure[]{\includegraphics[width=0.45\textwidth]{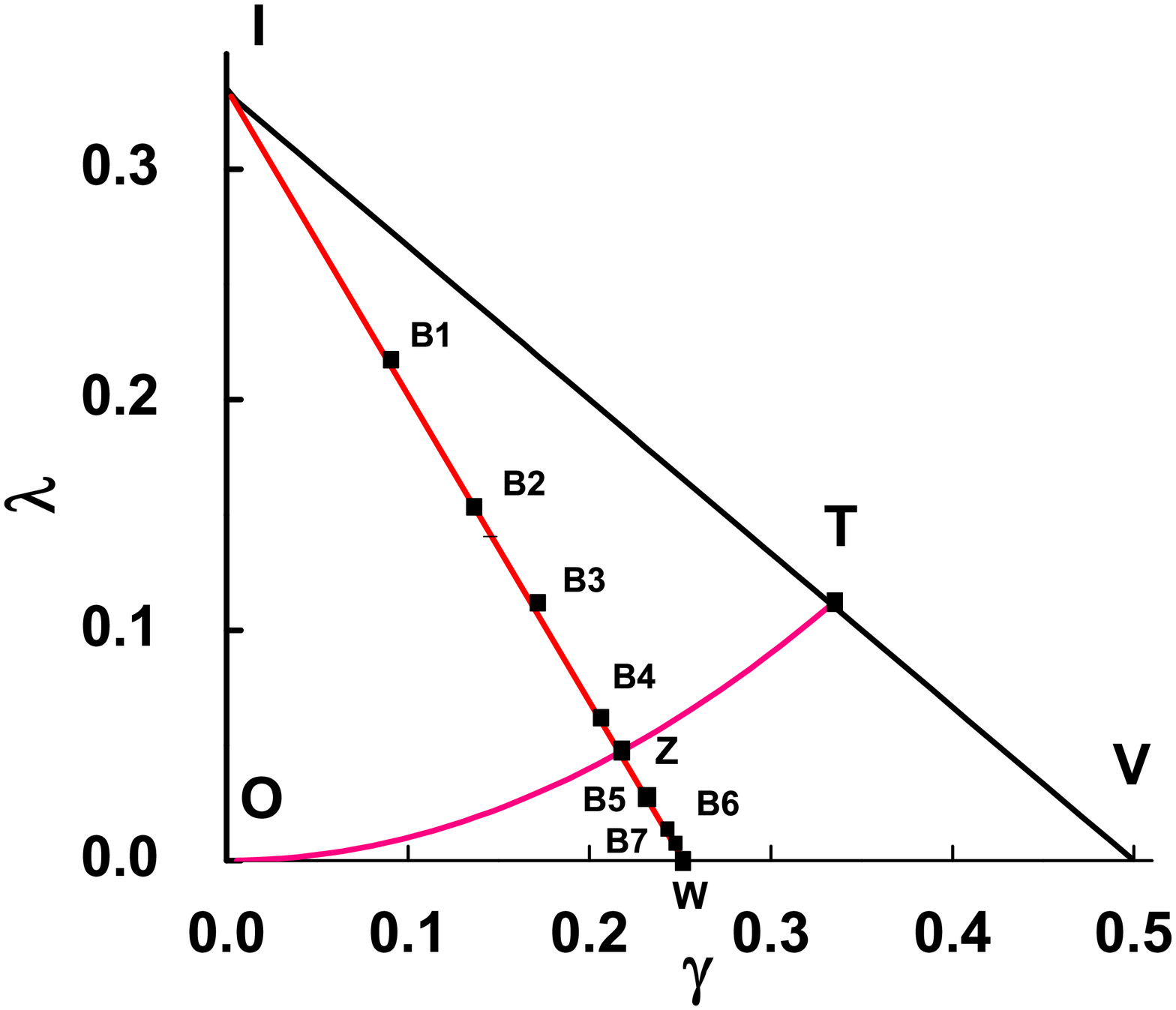}
\label{fig:1b}}
}
\caption{(a)The essential triangle, depicting the interior trajectory IZW
along which detailed simulations have been carried out. (b)Location of 
designated values of ($\gamma, \lambda$) along IW (also see Table.(\ref{tab:table1}))} 
\label{fig:1}
\end{figure}

The biaxial liquid crystal phase ($N_{B}$)which was predicted in the early 70's
\cite{freiser}continues to be elusive inspite of significant progress made in the
theoretical\cite{Straley} - \cite{Mukherjee09}, experimental\cite{Yu} - 
\cite{Severing} and computer simulation \cite{luck80} - \cite{matteis05B}
studies regarding the existence and properties of the phase. The 
recent mean field  theoretical (MFT) studies of a quadrupolar Hamiltonian 
model \cite{Sonnet} - \cite{matteis07} predict a universal mean field phase diagram 
for biaxial nematic along the boundary 
of a triangular parameter space OIV( see Fig.\ref{fig:1a}) 
 wherein the condensation of the biaxial phase could occur either from the uniaxial 
 ($N_{U}$) phase or directly from the isotropic phase ($I$). These 
 predictions, which were partly verified by Monte Carlo simulations,
 were found to be unsatisfactory in  the limit of vanishing biaxial-biaxial 
 interaction in the repulsive region for the Hamiltonian, thus requiring 
further study. 
     
     In this context, our recent WL simulations of the phase sequences along
 the boundary of the triangle OIV   
 \cite{kamala14,kamala15} suggested a qualitative modification 
of the MFT phase diagram as the Hamiltonian is driven to the partly repulsive regions.
The efficient entropic sampling technique employed \cite{Wang, Zhou, Jayasri09, kamala15} seeking otherwise inaccessible 
rare microstates, pointed to the existence of possible  hindering free energy 
barriers within the system resulting from the absence of stabilising long range 
order of one of the molecular axes. Keeping in view the crucial role played by 
the degree of cross-coupling between the 
uniaxial and biaxial tensorial components of the neighbouring molecules 
in the condensation of the biaxial phase, we present in this paper, the results 
of a similar detailed  simulation study  which was carried out along 
a segment IW in the interior of the essential triangle, where W is the 
midpoint of  OV (see Fig. \ref{fig:1}).
   
     We carried out a systematic 
simulation study using the entropic sampling technique (WL algorithm) along
the segment IW in order to obtain a generic phase diagram inside the 
essential triangle. The mean field Hamiltonian and the simulation model are 
presented in section II. The sampling technique and details of simulation 
are discussed in section III. The observations from the simulations are 
reported in section IV and the conclusions are discussed in section V.  

\section{Hamiltonian model}

The MF analysis \cite{Sonnet} - \cite{matteis07}, 
is based on the general quadrupolar orientational Hamiltonian, proposed by
 Straley \cite{Straley} and set in terms of tensors
\cite{Sonnet}. Accordingly, the interacting  biaxial molecules
are represented by two pairs of symmetric, traceless tensors 
($\bm{q}$, $\bm{b}$) and ($\bm{q^{'}}$, $\bm{b^{'}}$). Here $\bm{q}$ and 
$\bm{q^{'}}$ are uniaxial components about the unit molecular vectors $\bm{m}$ 
and $\bm{m^{'}}$, whereas $\bm{b}$ and $\bm{b^{'}}$ (orthogonal to
$\bm{q}$ and $\bm{q^{'}}$, respectively), are biaxial. These irreducible
components of the anisotropic parts of susceptibility tensor are 
represented in its eigen frame 
$(\bm{e},\bm{e_{\perp}},\bm{m})$ as
 
\begin{subequations}\label{s1}
\begin{align}
\bm{q} &:= \bm{m} \otimes \bm{m} - \frac{\bm{I}}{3} \\
\bm{b} &:= \bm{e} \otimes \bm{e} - \bm{e}_{\perp} \otimes \bm{e}_{\perp}
\end{align}
\end{subequations}
where $\bm{I}$ is the identity tensor. Similar representations hold for 
$\bm{q^{'}}$ and $\bm{b^{'}}$ in the eigen frame
 $(\bm{e^{'}}, \bm{e^{'}_{\perp}}, \bm{m^{'}})$. The interaction energy 
is written as

\begin{equation}
H=-U[\xi \, \bm{q} \cdot \bm{q}^{\, \prime}
+ \gamma(\bm{q} \cdot \bm{b}^{\, \prime} + \bm{q^}{\, \prime} \cdot \bm{b}) +
\lambda \, \bm{b} \cdot \bm{b}^{\, \prime}]
\label{eqn:w2}
\end{equation}
where $U$ is the scale of energy, $\xi$ = $\pm 1$, $ \gamma$ and
$\lambda$ are dimensionless interaction parameters, determining the 
relative importance of the uniaxial-biaxial coupling and 
biaxial-biaxial coupling interactions between the molecules, respectively.

\indent   Mean-field analysis of the Hamiltonian identifies a triangular 
region  OIV in the $(\gamma,\lambda)$ plane - called the essential triangle - 
representing the domain of stability into which any physical system
represented by Eqn.~\eqref{eqn:w2} can be mapped \cite{Bisi06, matteis07}
(see Fig.~ \ref{fig:1}). The line $C_{1}C_{3}$ is a tricritical line whereas
$C_{2}C_{3}$ is a triple line. The dispersion parabola $\lambda=\gamma^{2}$
 \cite{Luck75} traverses through the interior of the triangle, intersecting
 IV at the point T, called the Landau point. Region of the triangle above 
 the parabola corresponds to a Hamiltonian where all the terms are 
 attractive, while the region below is noted to be partly repulsive 
 \cite{Bisi06}.  MFT predicts a $I \rightarrow N_{U} \rightarrow N_{B}$ 
 phase sequence in the quadrangle $OC_{2} C_{3}V$ and a direct 
 $I\rightarrow N_{B}$ transition in the triangular region $IC_{2}C_{3}$.

 For simulation purposes, the general Hamiltonian in Eqn.~\eqref{eqn:w2} is 
 conveniently recast as a  biaxial mesogenic lattice model, where particles 
of $D_{2h}$ symmetry, represented by unit vectors $\bm{u}_{a}$, $\bm{v}_{b}$ 
on lattice sites a and b interact through a nearest-neighbour pair 
potential \cite{romano}

\begin{equation}
U= -\epsilon \lbrace G_{33} - 2\gamma(G_{11}-G_{22})+
\lambda[2(G_{11}+G_{22})-G_{33}]\rbrace.
\label{eqn:w6}
\end{equation}

Here  $f_{ab}$= ($\bm{v}_a$.$\bm{u}_b$),
 $G_{ab}$=$P_2$($f_{ab}$) with $P_{2}$ denoting the second Legendre 
polynomial. The constant $\epsilon$ (set to unity in
simulations) is  a positive quantity setting the reduced temperature
$\textit{T}^{'}=k_{B}\textit{T}/\epsilon $, where $\textit{T}$ is 
the absolute temperature of the system. 

\section{Details of Simulation}
The Wang-Landau (WL) sampling \cite{Wang} is a flat histogram technique 
designed to overcome energy barriers encountered, for example,
 near first order transitions, by facilitating
a uniform random walk along the energy ($\textit{E}$) axis through an 
appropriate algorithmic guidance. The sampling, originally developed for  
Hamiltonian models involving random walks in discrete configurational 
space, continues to be applied to various problems in statistical
physics \cite{ Landau1, Murthy1}, polymer and  protein studies 
\cite{Rathore03, Seaton10, Priya11} and 
is being developed for more robust applications for continuous systems
\cite{Poulain,Sinha09,Raj,Yang13, Vogel13, Katie14, Xie14} and self 
assembly \cite{Landau13}. The proposed algorithm was modified \cite{jayasri} 
to suit lattice models like the Lebwohl-Lasher interaction \cite{LL}, 
allowing for continuous variation of molecular orientations. It was 
subsequently augmented with the so-called $\textit{frontier}$ sampling 
technique \cite{Zhou, Jayasri09} to simulate more complex systems like 
the biaxial medium. The WL sampling is based on effecting a convergence 
of an initial distribution over energy $\textit{E}$ to the density of 
states (DoS) $g(E)$ of the system iteratively. Frontier sampling technique
is an algorithmic guidance, provided in addition to the WL routine, by 
which the system is constrained to visit and sample from low entropic 
regions. 

The simulations are performed on a cubic lattice of dimensions 
($L \times L \times L, L=15 and 20$) with periodic boundary conditions.
The biaxial liquid crystal molecule on each lattice site 
interacts with the nearest neighbours based on the 
potential in Eqn.(\ref{eqn:w6}). The uniaxial - biaxial coupling 
coefficient $\gamma$ on IW is  half of the the value on the 
diagonal IV, for identical $\lambda$ values. We denote the arclength
of the path OIW as $\lambda^{'}$, given by $\lambda^{'} = \lambda$ on 
segment OI, and 
$$
\lambda^{'}= \dfrac{1}{3}(1 + 5\gamma)
$$
where
$$
\gamma= \dfrac{(1-3\lambda)}{4}
$$
on the segment IW.\\

The parameters $\gamma$ and $\lambda$ were chosen such that we traverse 
along the path  $IW$ which amounts to varying the 
arclength $ \lambda^{'}$ from ~0.33 to ~0.747.

\begin{table}[h!]
\caption{Coordinates of points  B1-B7 and Z along the segment IW of the 
essential triangle (Fig.2)}
 \begin{center}
 \renewcommand{\arraystretch}{1.2}
  \begin{tabular}{ | c | c | c | c |}
      \hline
     Point &  $\gamma$   &  $\lambda$  &  $\lambda^{'}$    \\ \hline
     B1 & 0.0859 & 0.1719  & 0.4766 \\[2ex] \hline
     B2 &0.1405  & 0.14658 & 0.5666 \\ [2ex] \hline
     B3 &0.1663  & 0.1116  & 0.6105  \\ [2ex]  \hline
     B4 & 0.2045 & 0.0606  & 0.674  \\ [2ex]  \hline
     Z  & 0.2149 & 0.0467  & 0.691  \\ [2ex]  \hline
     B5 & 0.2253 & 0.0328  & 0.709 \\ [2ex]  \hline
     B6 & 0.2440 & 0.0079  & 0.740 \\  [2ex]  \hline
     B7 & 0.2482 & 0.0024  & 0.747\\ [2ex]
    \hline
    \end{tabular}
    \label{tab:table1}
\end{center}
\end{table}

The Table~\ref{tab:table1} lists the values of 
 ($\gamma, \lambda$) and corresponding arc lengths $\lambda^{'}$ 
 at some designated points for ready reference. Fig.~\ref{fig:1b} shows the 
 location of these points schematically inside the 
 essential triangle. Simulations were carried 
out for nearly 40 points on IW using modified Wang-Landau 
 (WL) algorithm augmented by frontier sampling. At each  value, 
 the g(E) which is the  estimate 
 of the density of states was obtained and an entropic ensemble of 
 $10^7$ microstates  was generated,  by making an effectively 
 uniform random walk in  energy space guided by the DoS. Equilibrium 
 ensembles at any desired (reduced) temperature ($T^{'}$) are consequently
 extracted by a suitable reweighting procedure  
 \cite{Zhou, Swensden} and the average values of physical properties are
calculated. The representative free energy $\textsl{F}$, as a function of the energy 
of the system, as well as of the two dominant order parameters (uniaxial 
and biaxial orders) is computed from the DoS and the 
microcanonical energy, - both available as a function of bin number
in the entropic ensemble. 

    The physical parameters of interest in this system, calculated at 
each $\lambda^{'}$, are the average energy $<E>$, specific heat $<C_{v}>$, 
energy cumulant $V_{4}$ (= $1-<E^{4}>/(3<E^{2}>^{2})$) which is a measure 
of the kurtosis \cite{Binder}, the four order parameters of the phase 
calculated according to \cite{Biscarini95,Robert} and their susceptibilities. 
These are the uniaxial order $<R^{2}_{00}>$ (along the primary director), 
the phase biaxiality $<R^{2}_{20}>$, and the molecular contribution to 
the biaxiality of the medium  $<R^{2}_{22}>$, and the contribution
to uniaxial order from the molecular minor axes $<R^{2}_{02}>$. 
 
 The averages are computed at a temperature resolution of 0.002 units in 
 the temperature range [0.05, 2.05]. Statistical errors in different observables 
 are estimated over ensembles comprising a minimum of $5 \times 10^{5}$ 
 microstates, and these are compared with several such equilibrium ensembles 
 at the same  $(\gamma,\lambda)$ value, but initiating the random walk 
 from different arbitrary points in the configuration space. We find the 
 relative errors in energies are 1 in $10^{5}$, while those in the estimation 
 of the order parameters are 1 in $10^{4}$. 
 
 \section{Results and Discussion}
WL simulations were carried out at 40 values of $\lambda^{'}$ on the 
segment IW, where the arc length $\lambda^{'}$ ranges from 
0.33 to 0.75. Temperature variation of the specific heat, and the two
order parameters ($R^{2}_{00}$ and $R^{2}_{22}$) in different ranges of 
$\lambda^{'}$, covering the segment IZ (in the attractive region of the
interaction Hamiltonian), are presented in  
Figs.~\ref{fig:2} - \ref{fig:4}. The corresponding data along the segment 
ZW (in the repulsive region) is presented in Fig.\ref{fig:14}.
  
\subsection{Segment IZ: Range of $\lambda^{'}$ = (0.33 - 0.691)}

 \begin{figure}[htp]
\centering{
\subfigure[]{\includegraphics[width=0.45\textwidth]{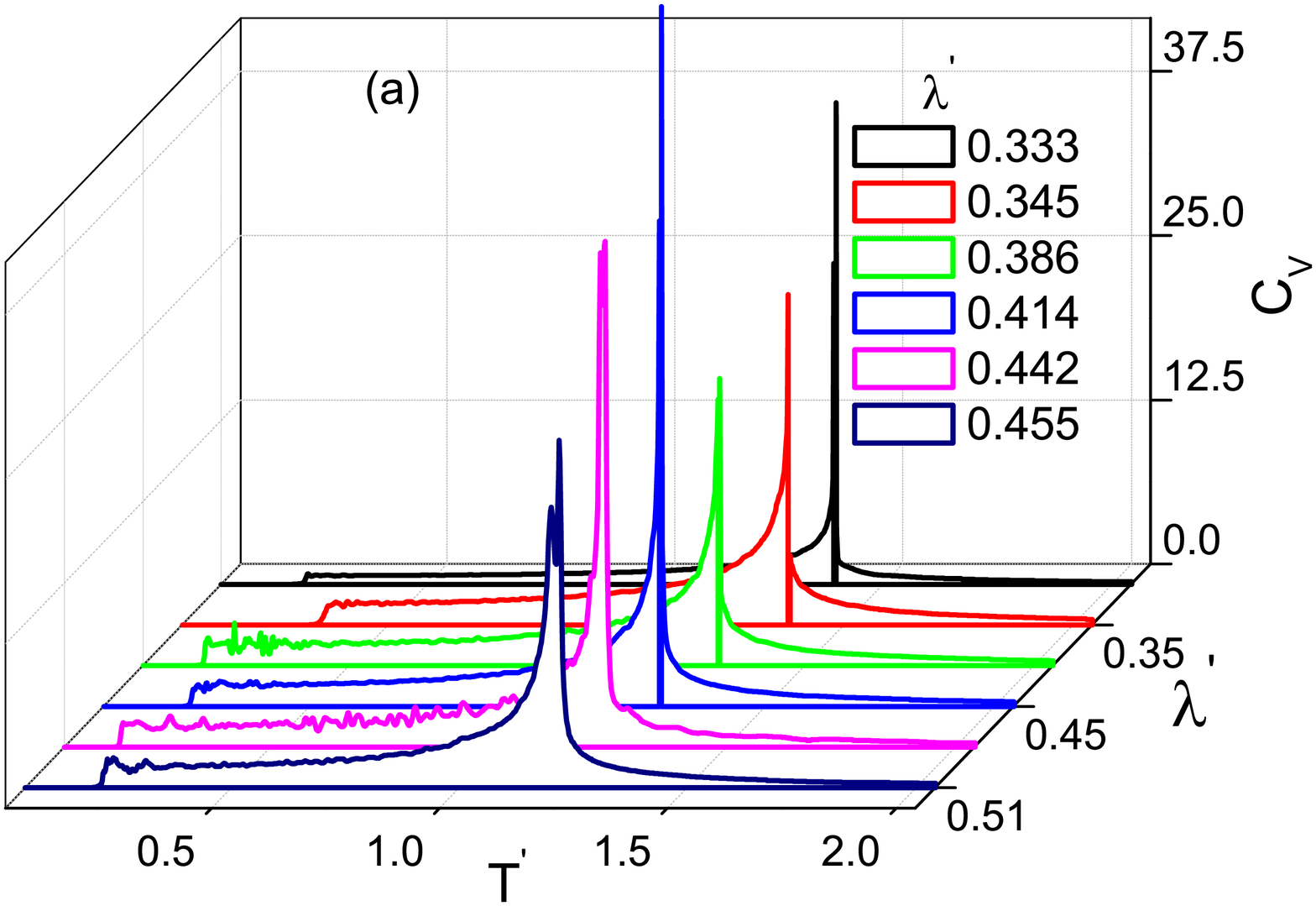}
\label{fig:2a}}
\subfigure[]{\includegraphics[width=0.45\textwidth]{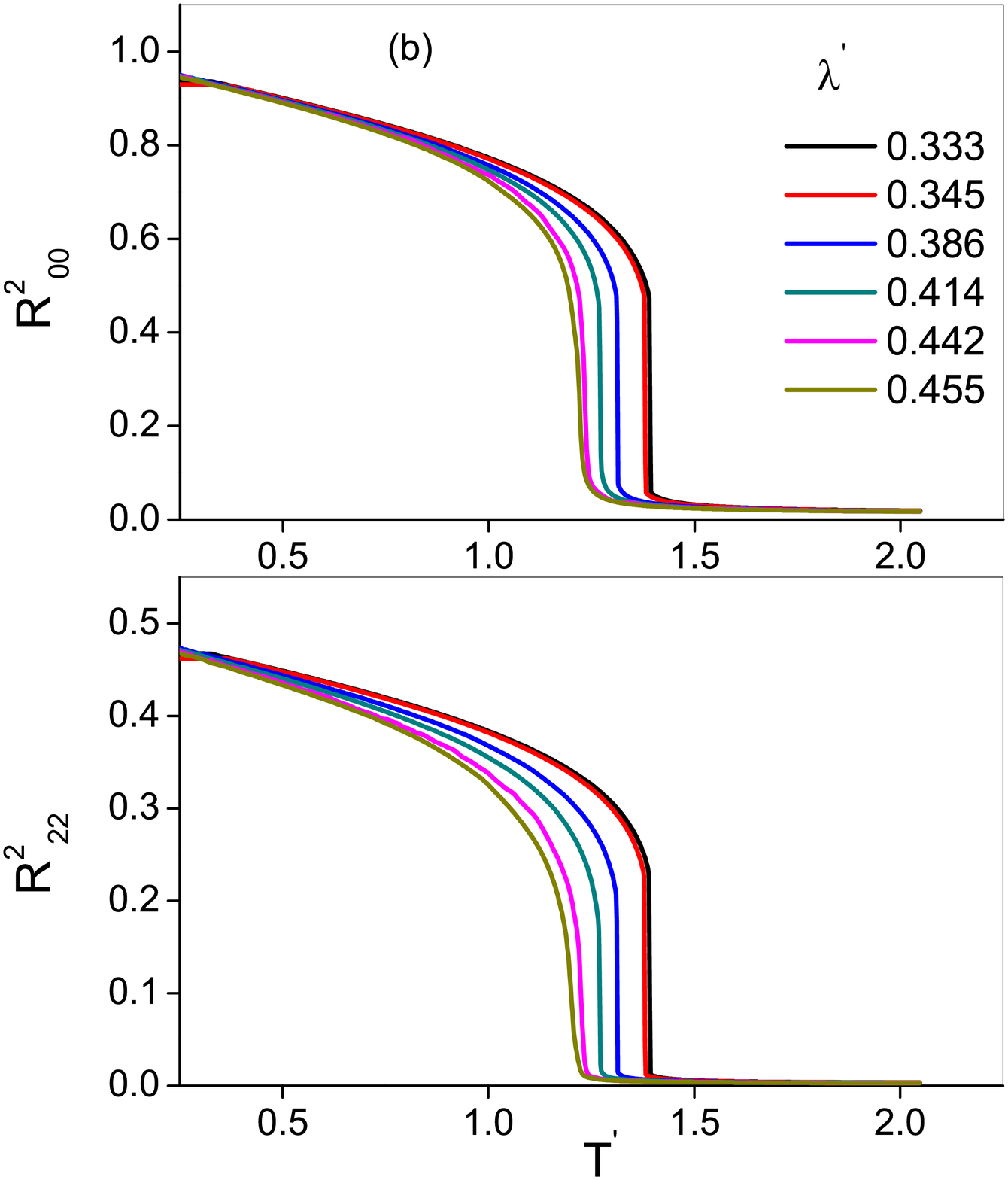}
\label{fig:2b}}
}
\caption{Comparison of (a) specific heat (b) order profiles for 
values of $\lambda^{'}$ from 0.33 to 0.455 } 
\label{fig:2}
\end{figure}
\begin{figure}[htp]
\centering{
\subfigure[]{\includegraphics[width=0.45\textwidth]{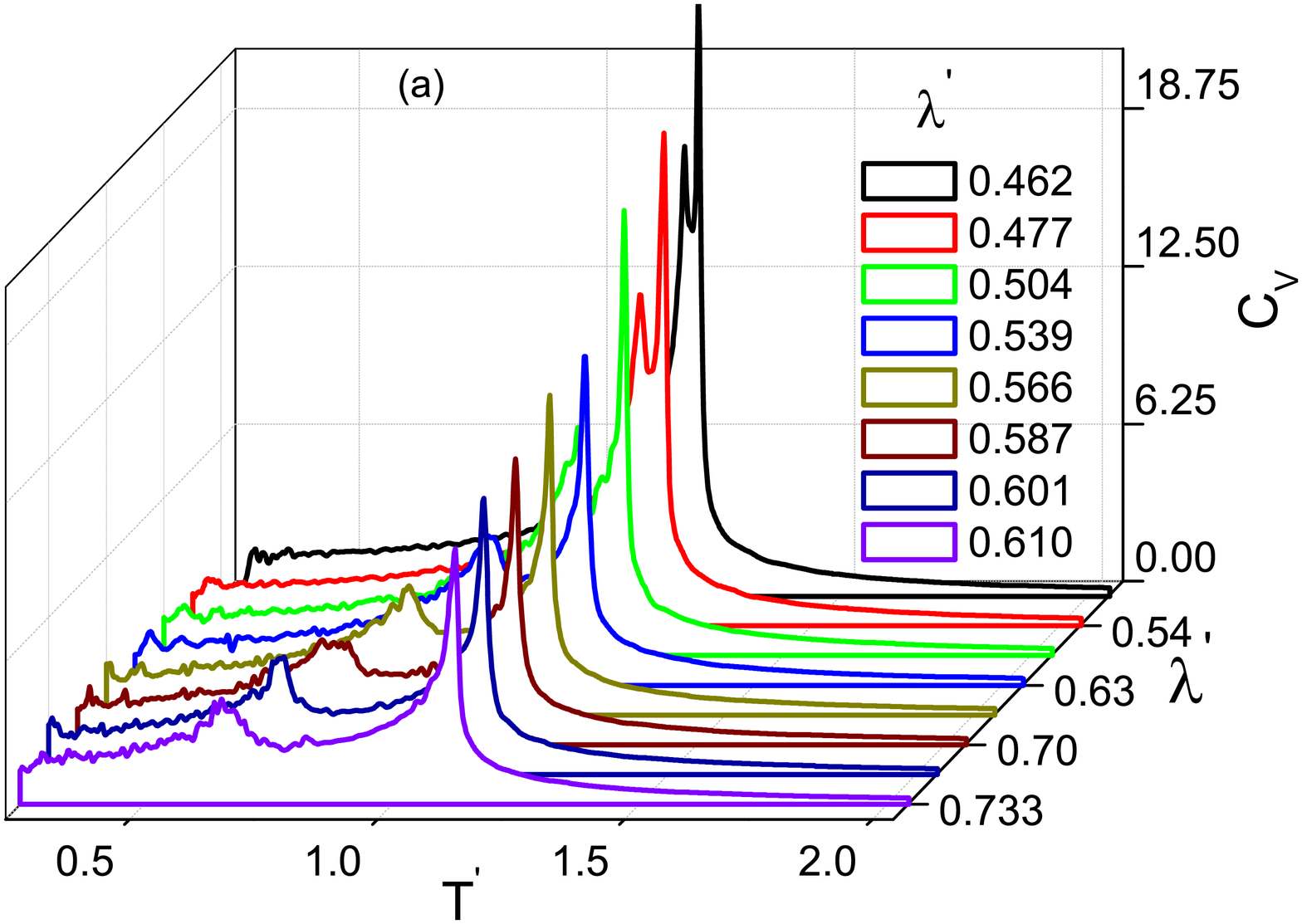}
\label{fig:3a}}
\subfigure[]{\includegraphics[width=0.45\textwidth]{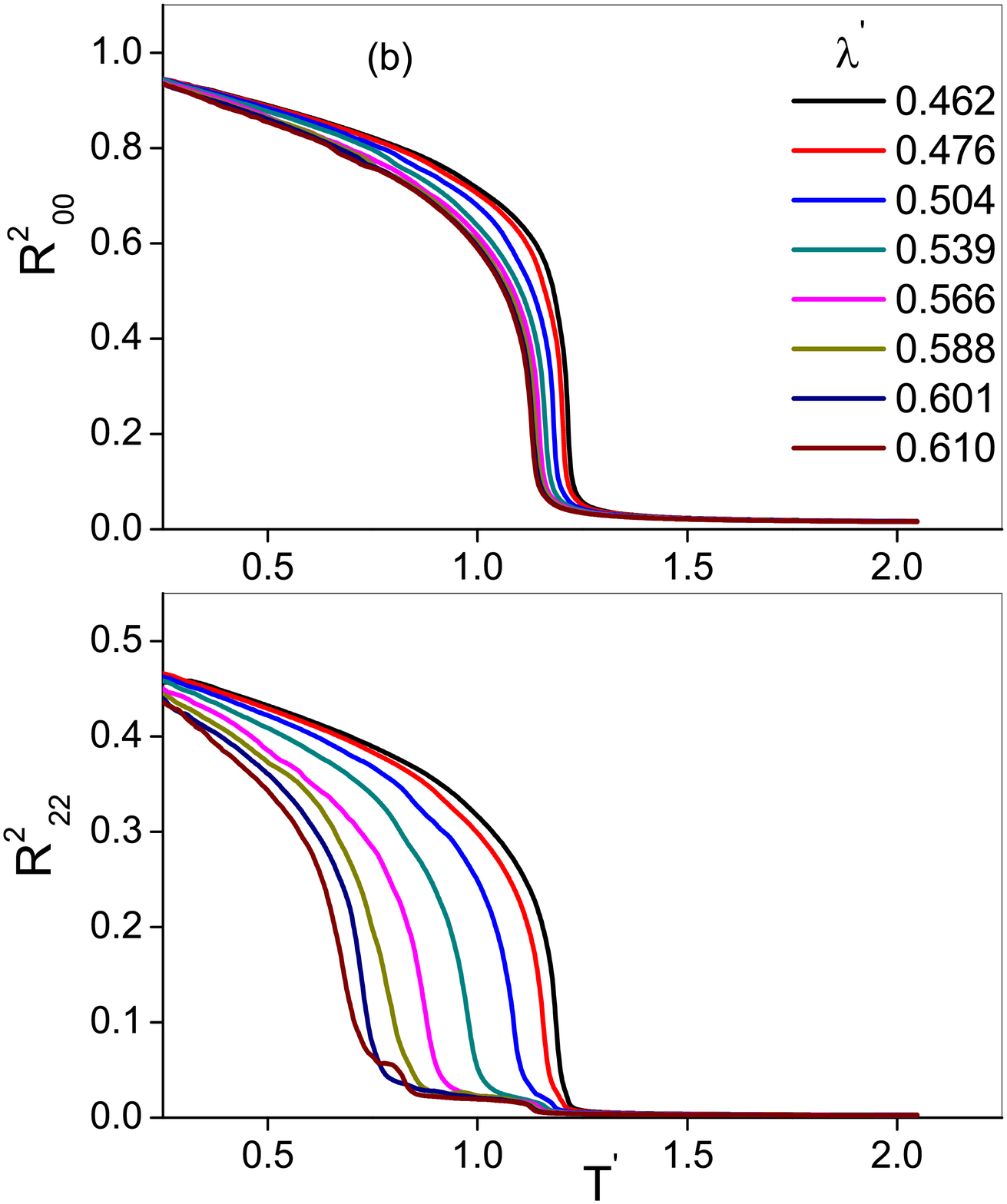}
\label{fig:3b}}
}
\caption{Comparison of (a) Specific heat (b) order parameter profiles for 
 values of $\lambda^{'}$ from 0.463 to 0.610 } 
\label{fig:3}
\end{figure}

It is noted from Fig.~\ref{fig:2a} that for all values of 
$\lambda^{'}$  in the range 0.33 - 0.455, a single transition peak is
 observed in the specific heat profile. As the biaxial system is cooled 
 from the high temperature isotropic phase, a direct $I - N_{B}$ transition 
 takes place and the order profiles in Fig.~\ref{fig:2b} reflect the behaviour. 
Fig.~\ref{fig:3a} depicts the splitting of this transition 
into two transitions for higher values of  $\lambda^{'}$ in the range 
0.462 to 0.610. It is observed that lower transition temperature $T_{2}$ 
   is progressively depressed with increase in $\lambda^{'}$ value, as compared 
   to the higher transition temperature peak at $T_{1}$.
The variation of the order profiles in this 
region, shown in Fig.~\ref{fig:3b},  reveals an intervening phase which is 
not strictly  uniaxial since the system exhibits  a low value of $R^{2}_{22}$
at the onset of the transition. By performing simulations at different 
sizes (L=10, 15, 20), the possibility that this could be a finite size effect
is ruled out. (It may be noted that for these system sizes a pure uniaxial
phase condenses on the $\lambda$-axis). The notable difference in the 
case of path IW, relative to IV, is that the degree of biaxiality (value 
of $R^{2}_{22}$) remains fairly independent of temperature, and the degree 
is the same for all subsequent values of $\lambda^{'}$ beyond this threshold,
 until interrupted by a second low temperature transition leading to 
an onset of appreciable biaxial order (Fig.\ref{fig:3}).

 \begin{figure}[htp]
\centering{
\subfigure[]{\includegraphics[width=0.45\textwidth]{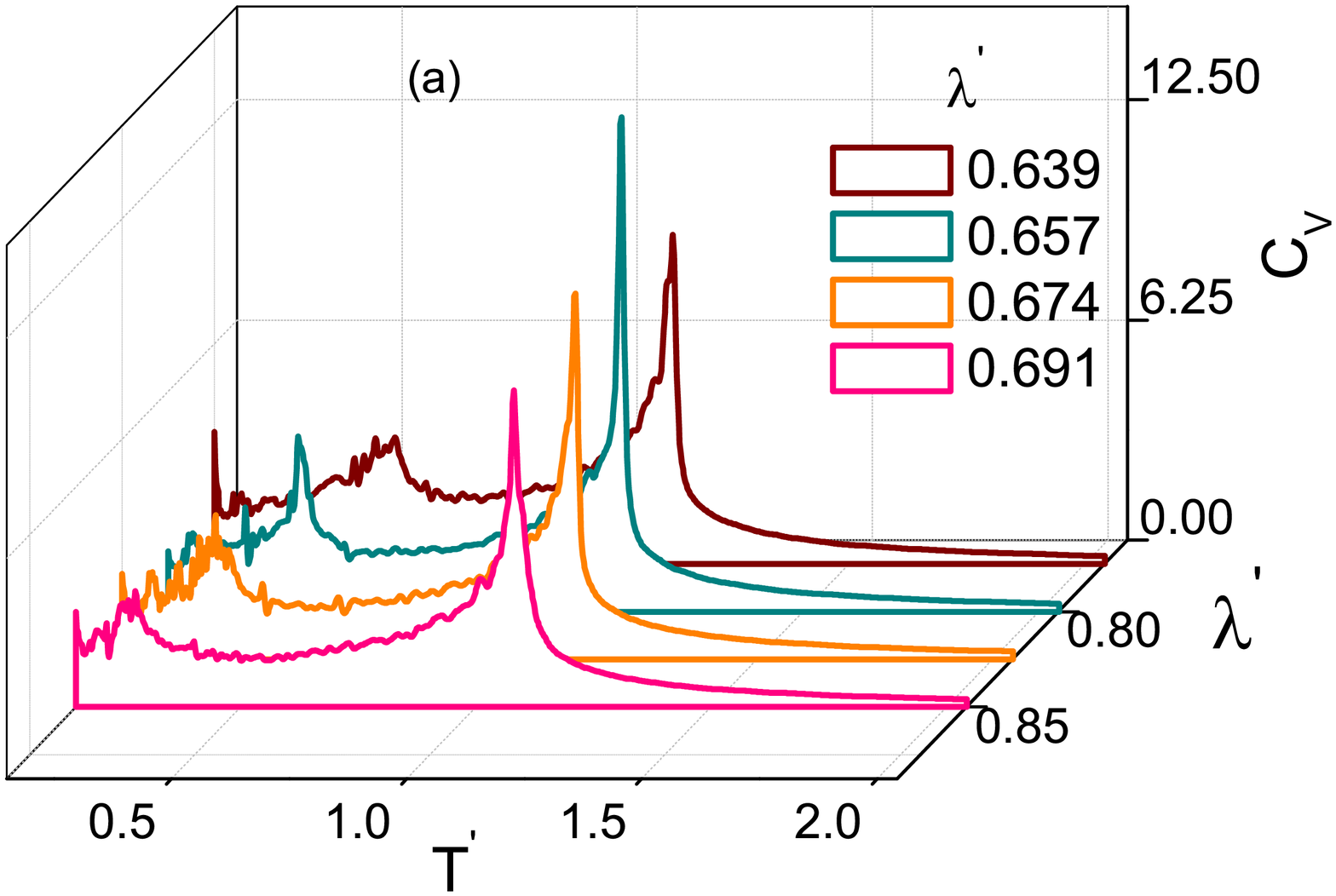}
\label{fig:4a}}
\subfigure[]
{\includegraphics[width=0.45\textwidth]{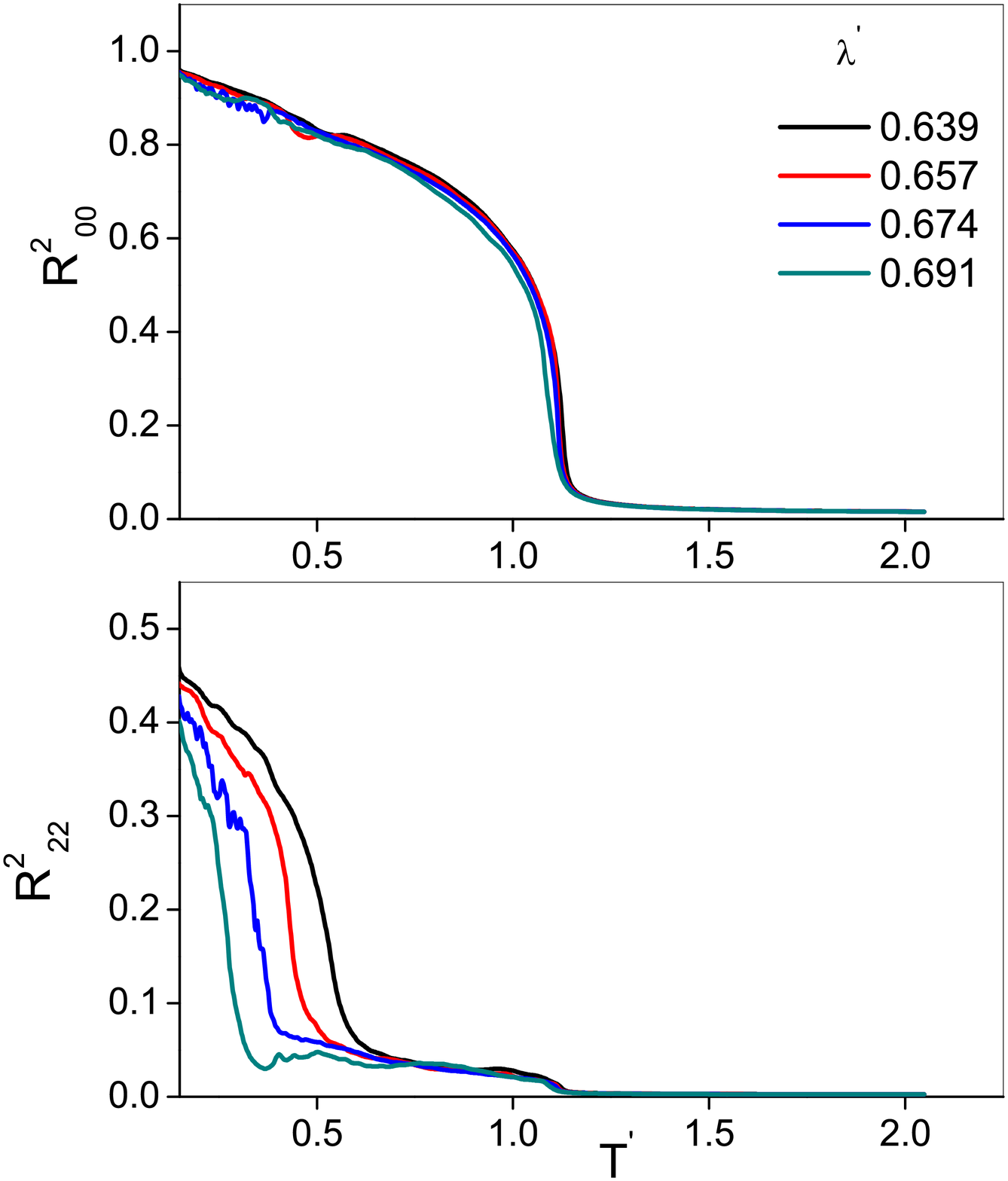}
\label{fig:4b}}
}
\caption{Comparison of (a) specific heat (b) order parameter profiles for 
values of $\lambda^{'}$ from  0.639 to 0.691 } 
\label{fig:4}
\end{figure}

\indent The results of simulation  for $\lambda^{'}$ values in the range  
0.639 - 0.691 are depicted in Fig.~\ref{fig:4}. Though these points lie 
very close to the parabola, they are still in the attractive region for the
interaction Hamiltonian. It is observed from Fig.~\ref{fig:4a} that the
second specific heat peak shifts  progressively to lower temperatures as the 
value of $\lambda^{'}$ increases. Corresponding variations in the order 
parameters, shown in Fig.\ref{fig:4b} confirm the shift of the second 
transition temperature to lower values. However it is observed that 
the equilibrium averages of order parameters 
 in this region of $\lambda^{'}$  are not as smooth, and show discernible
 fluctuations in the low temperature biaxial phase.
 
  It is interesting to  note  that the intermediate phase persists to 
 have a small degree of  biaxial order ($\sim0.05$) which (a) is not a 
 finite size effect; (b) is fairly  independent of temperatures within the 
 liquid crystal phase; and (c) does not depend on the values of 
 $\lambda^{'}$.  This phase with temperature dependence typical of 
 uniaxial order, but having a small and constant biaxial symmetry 
 ($\leq 0.05)$) , is designated as $N_{U^{'}}$ phase in our notation. 
 On subsequent lowering of temperature from this phase, 
the biaxial order increases rapidly at the second 
transition at $T_{2}$ and the lower temperature phase has  
macroscopically observable biaxial order, for all values of $\lambda^{'}$.

\begin{figure}[htp]
\centering
\includegraphics[width=0.45\textwidth]{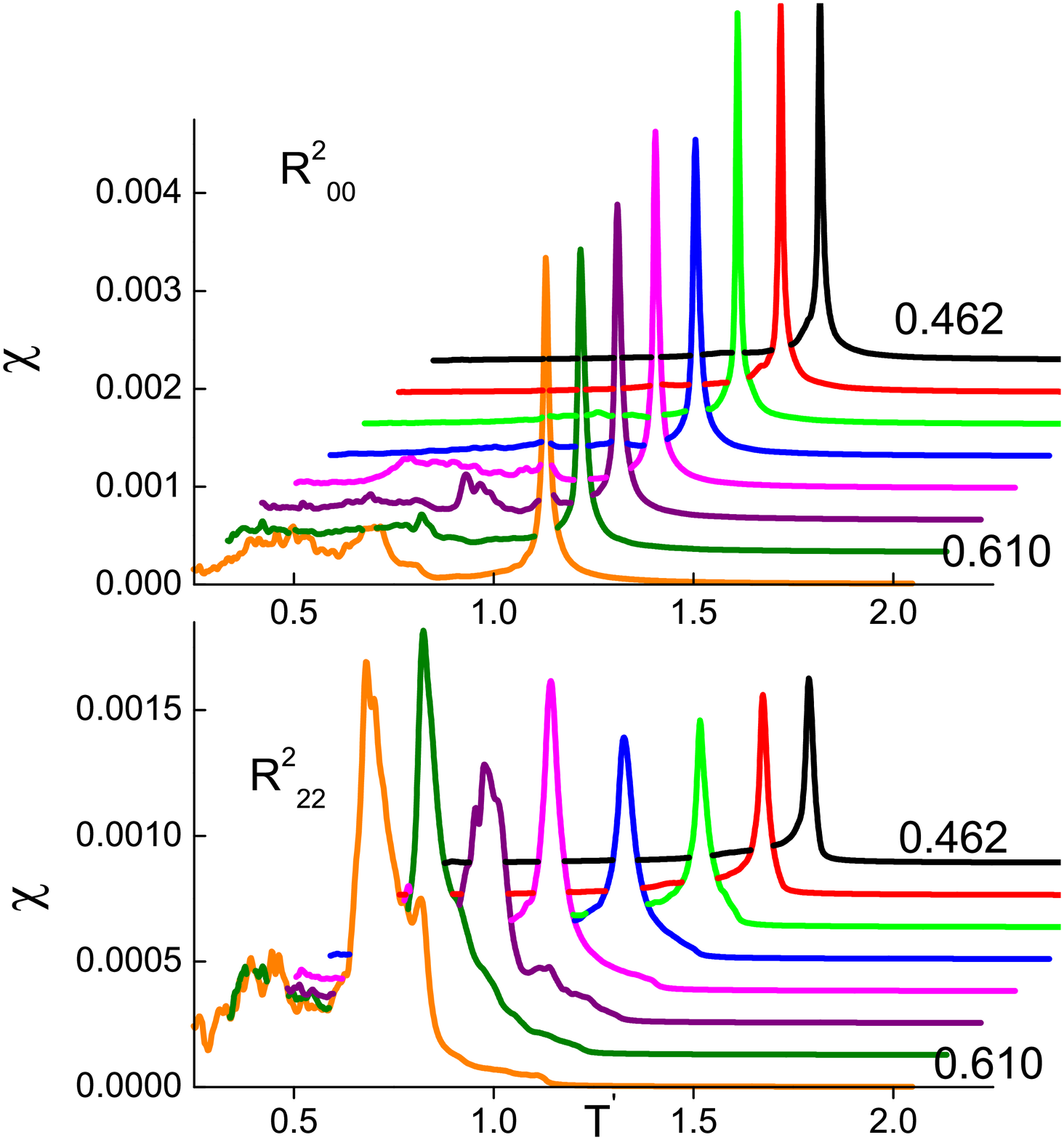}
\caption{Susceptibilities ($\chi^{\ '}$s) of the order parameters  for 
values of $\lambda^{'}$ : 0.462 - 0.610} 
\label{fig:5}
\end{figure}
   
 The susceptibility profiles of the order 
parameter in this region are depicted in Fig.~\ref{fig:4}. It is 
observed that the $R^{2}_{22}$ susceptibility starts increasing in the 
intermediate phase before showing a peak at the low temperature
transition at $T_{2}$. 

 \begin{figure}[htp]
\centering
\includegraphics[width=0.45\textwidth]{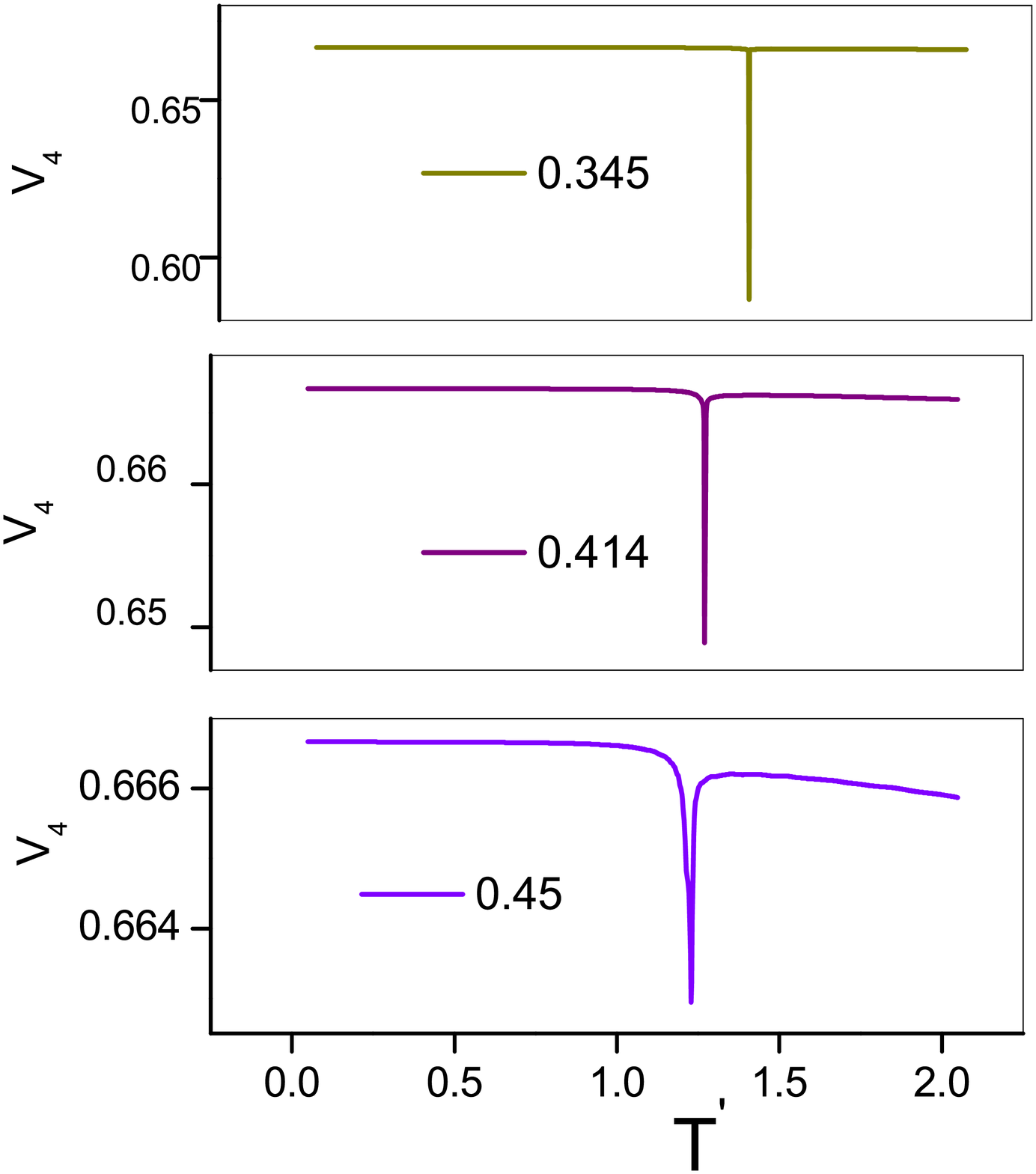}
\caption{energy cumulant $V_{4}$  for values of $\lambda^{'}$: 0.345 - 0.45} 
\label{fig:6}
\end{figure}

\begin{figure}[htp]
\centering
\includegraphics[scale=0.5]{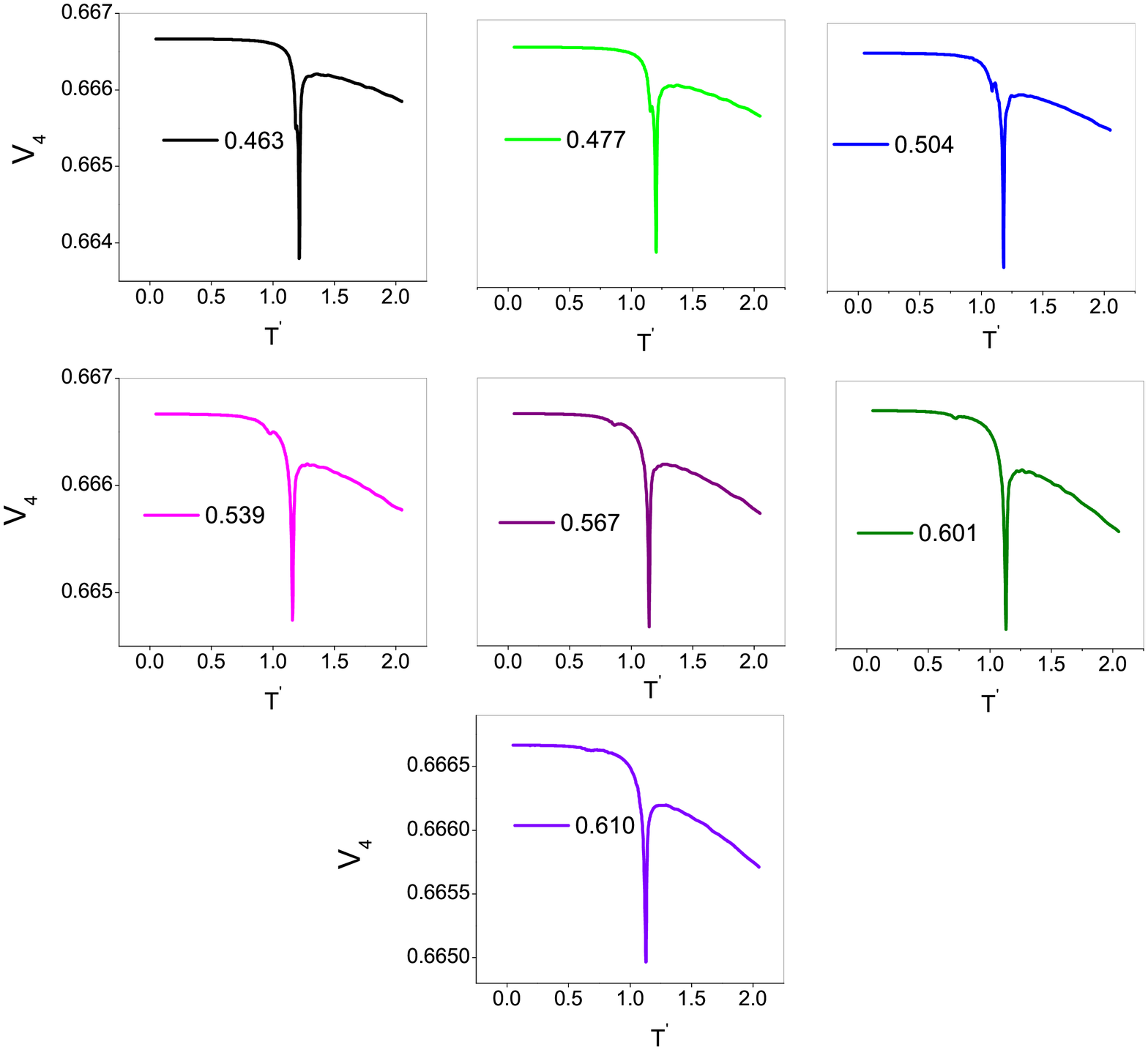}
\caption{Energy cumulant $V_{4}$ for values of $\lambda^{'}$ :0.463 - 0.610} 
\label{fig:7}
\end{figure}

\begin{figure}
\centering
\includegraphics[scale=0.4]{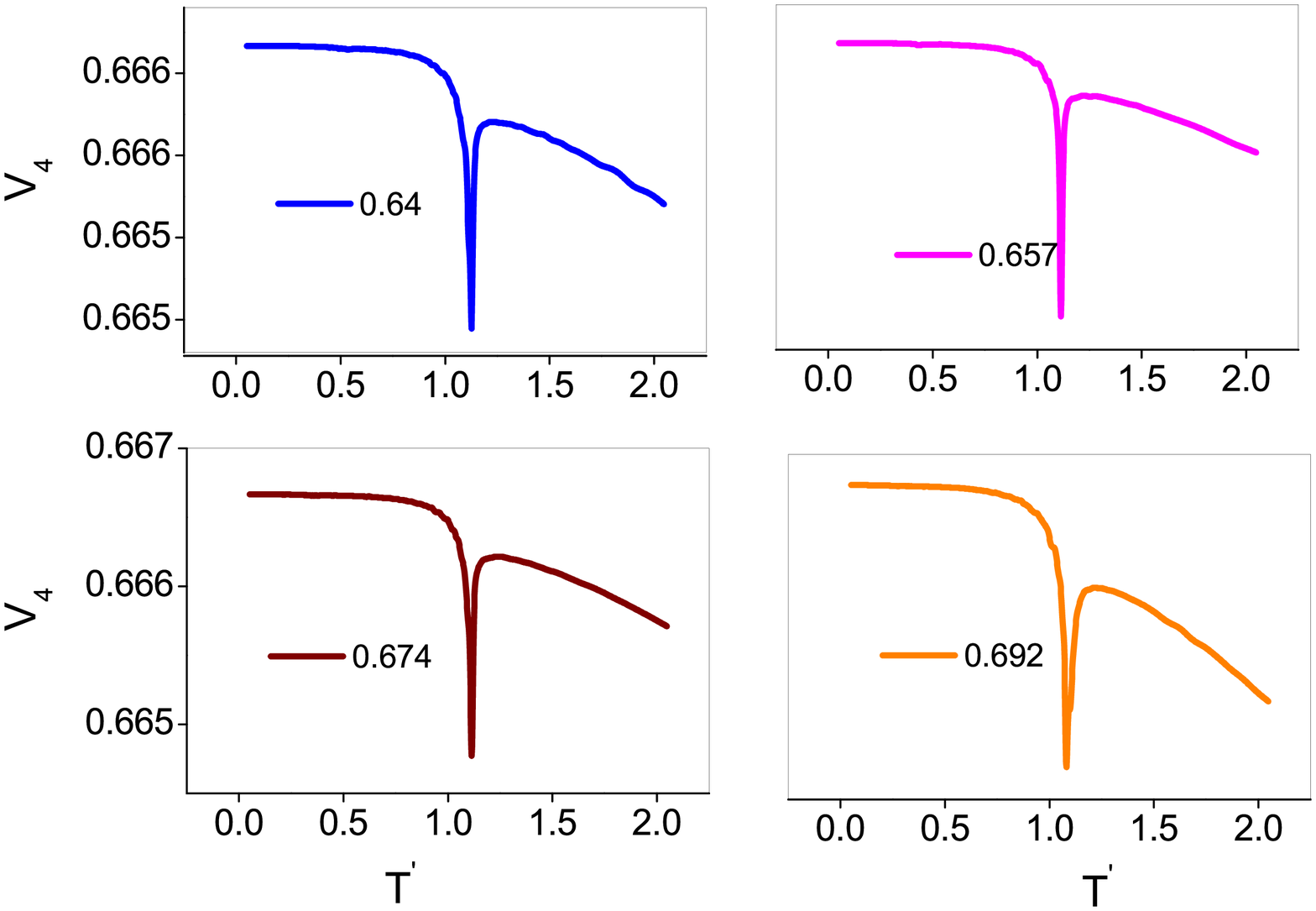}
\caption{energy cumulant $V_{4}$ for values of  $\lambda^{'}$ : 0.64 - 0.691} 
\label{fig:8}
\end{figure}

The fourth order energy cumulant ($V_{4}$) data obtained along the path OIW 
are shown in Figs.~\ref{fig:6} - \ref{fig:8}. It is observed that
the  $I-N_{B}$ transition remains strongly first order for values of $\lambda^{'}$
from 0.345  to 0.45. In the range of $\lambda^{'}$ from  0.463 to 0.691 ( i.e upto 
the point Z in Fig.\ref{fig:1}), the high 
temperature transition at $T_{1}$  from the isotropic phase (I) to the ordered
 $N_{U^{'}}$ phase shows a first order nature. Subsequently, the low temperature 
$N_{U^{'}} - N_{B}$ transition seems to change gradually from first order
to continuous nature, as seen from $V_{4}$ profiles in 
Figs.\ref{fig:6} and \ref{fig:8}. \\

\begin{figure}
\centering
\includegraphics[scale=0.5]{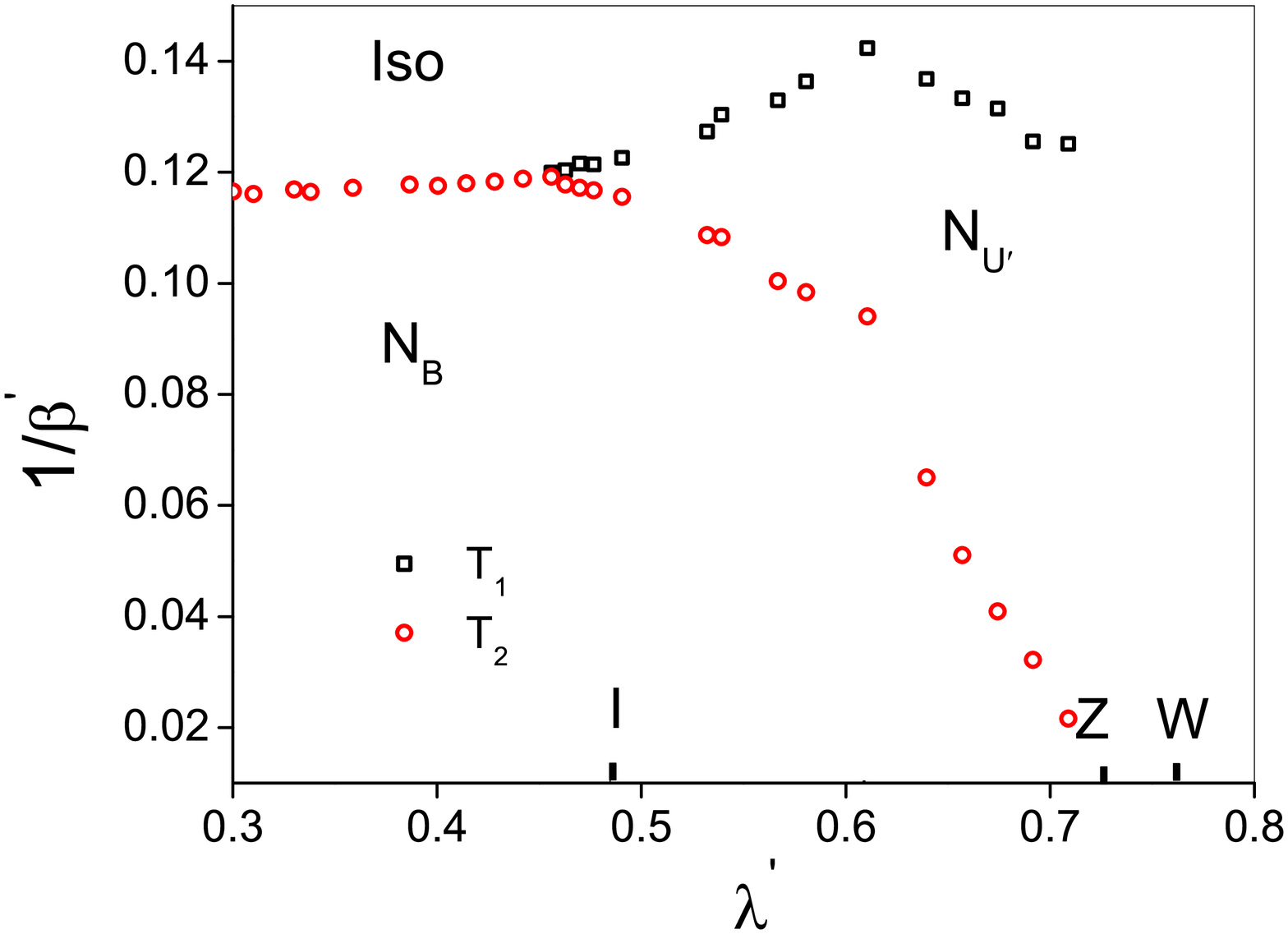}
\caption{Phase diagram inside the essential triangle along path  IW} 
\label{fig:9}
\end{figure}

An analysis of the above simulation data leads to the proposal of a phase 
diagram along the path IW, shown in Fig.\ref{fig:9}. We could report 
the $L=20$ data  only upto the value 
 $\lambda^{'} = 0.709$, as beyond this value (which falls into the partly 
 repulsive region under the parabola) the computational times for the 
 convergence of DoS are impractical.
 We observe from the temperature variation of order parameters that
 the growth of biaxial order appears to be progressively inhibited 
 as $\lambda^{'}$ value increases within the attractive region, and enters 
 the party repulsive region on crossing the parabola at  the point Z.  
The free energy profiles, plotted as a function of energy and 
order parameters (computed from the DOS data), reflect the rationale
for the impediments for the growth
of the biaxial order as the base of the triangle OIW is reached.
\begin{figure}
\centering{
\subfigure[]{\includegraphics[width=0.45\textwidth]{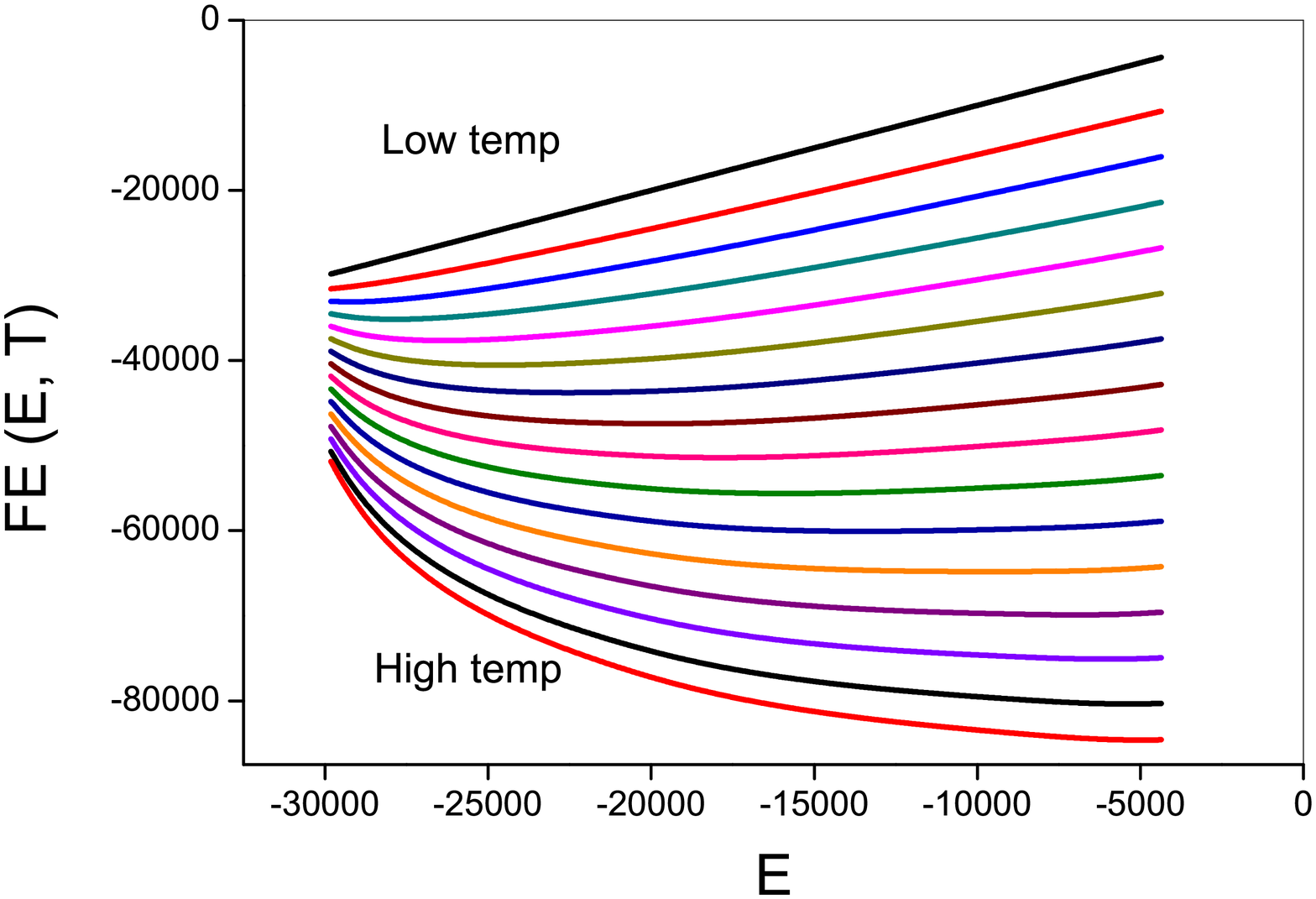}
\label{fig:10a}}
\subfigure[]{\includegraphics[width=0.45\textwidth]{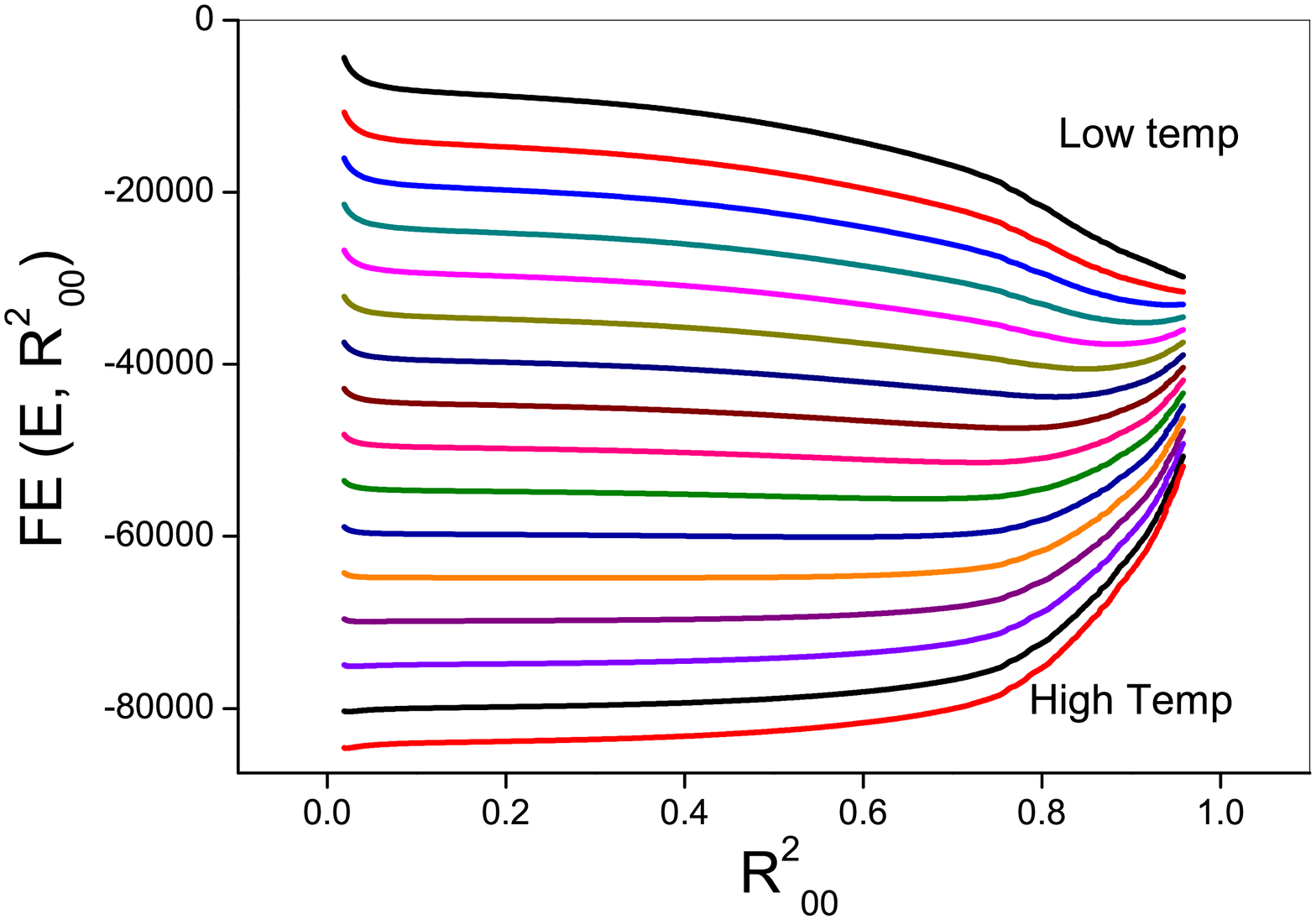}
\label{fig:10b}}
\subfigure[]{\includegraphics[width=0.45\textwidth]{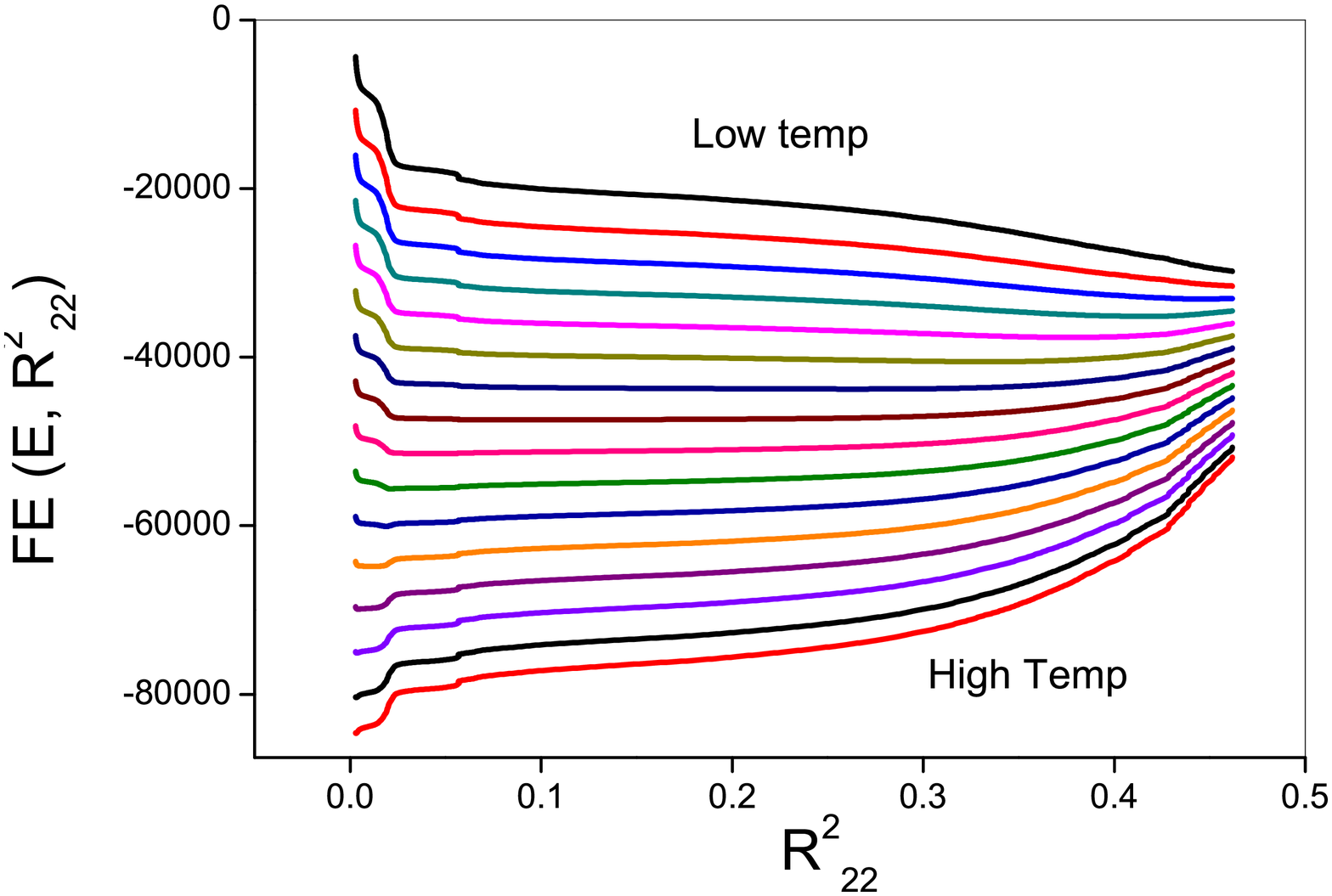}
\label{fig:10c}}
}
\caption{Representative free energy plotted as a function of (a) energy 
(b)$R^{2} _{00}$ (c) $R^{2}_{22}$  at the point B3 ($\lambda^{'}$ = 0.610
) } 
\label{fig:10}
\end{figure}

\begin{figure}
\centering{
\subfigure[]{\includegraphics[width=0.45\textwidth]{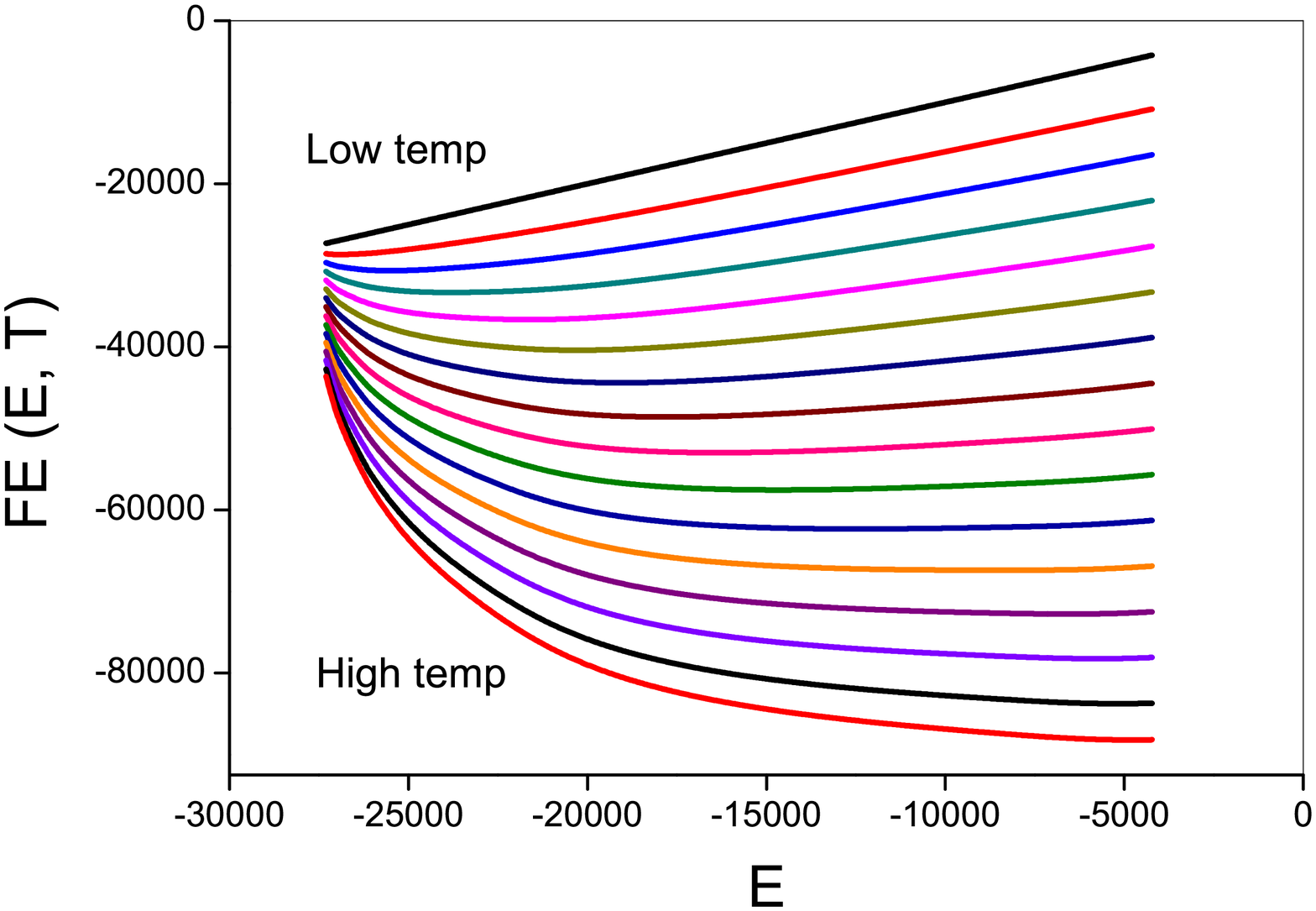}
\label{fig:11a}}
\subfigure[]{\includegraphics[width=0.45\textwidth]{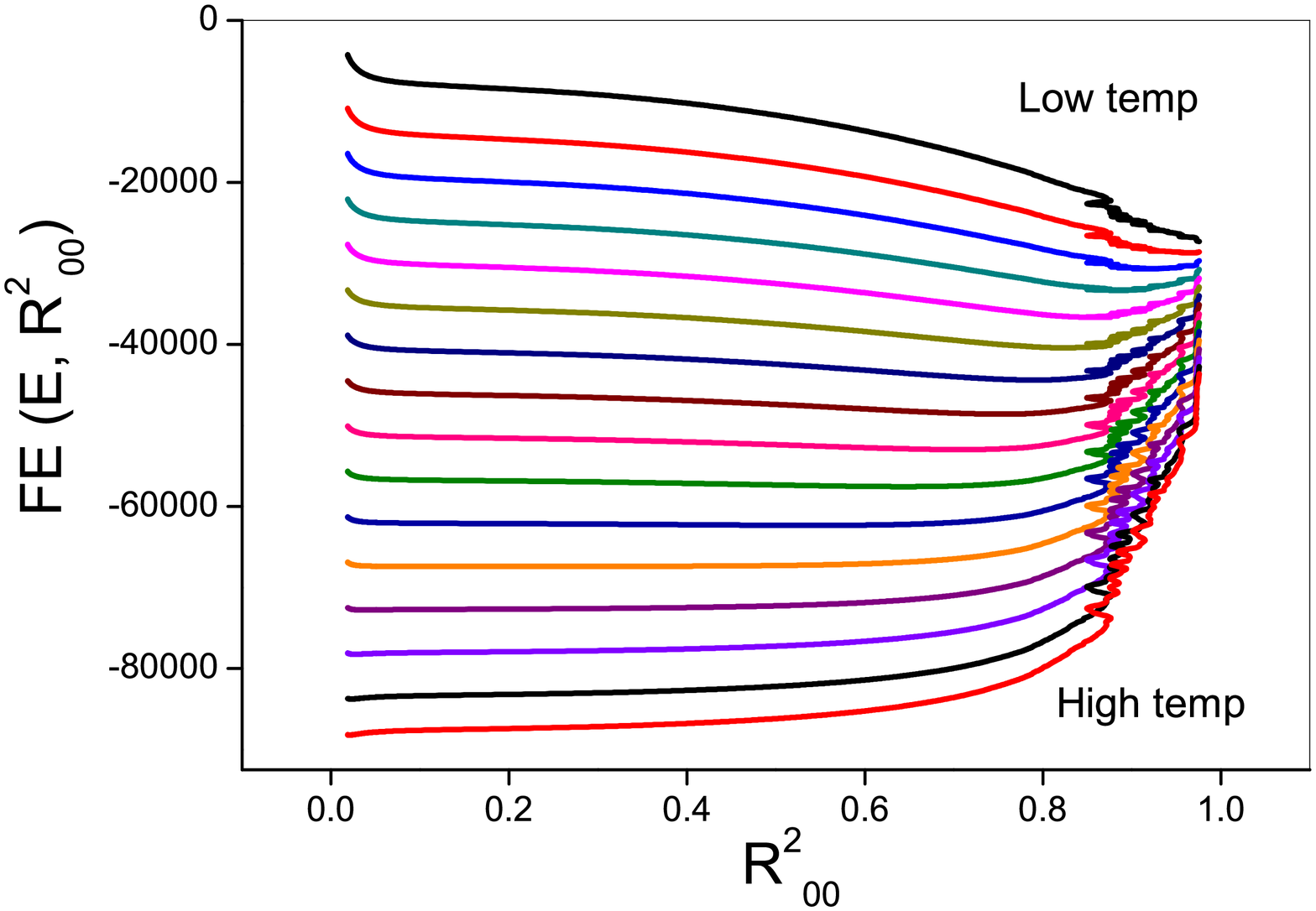}
\label{fig:11b}}
\subfigure[]{\includegraphics[width=0.45\textwidth]{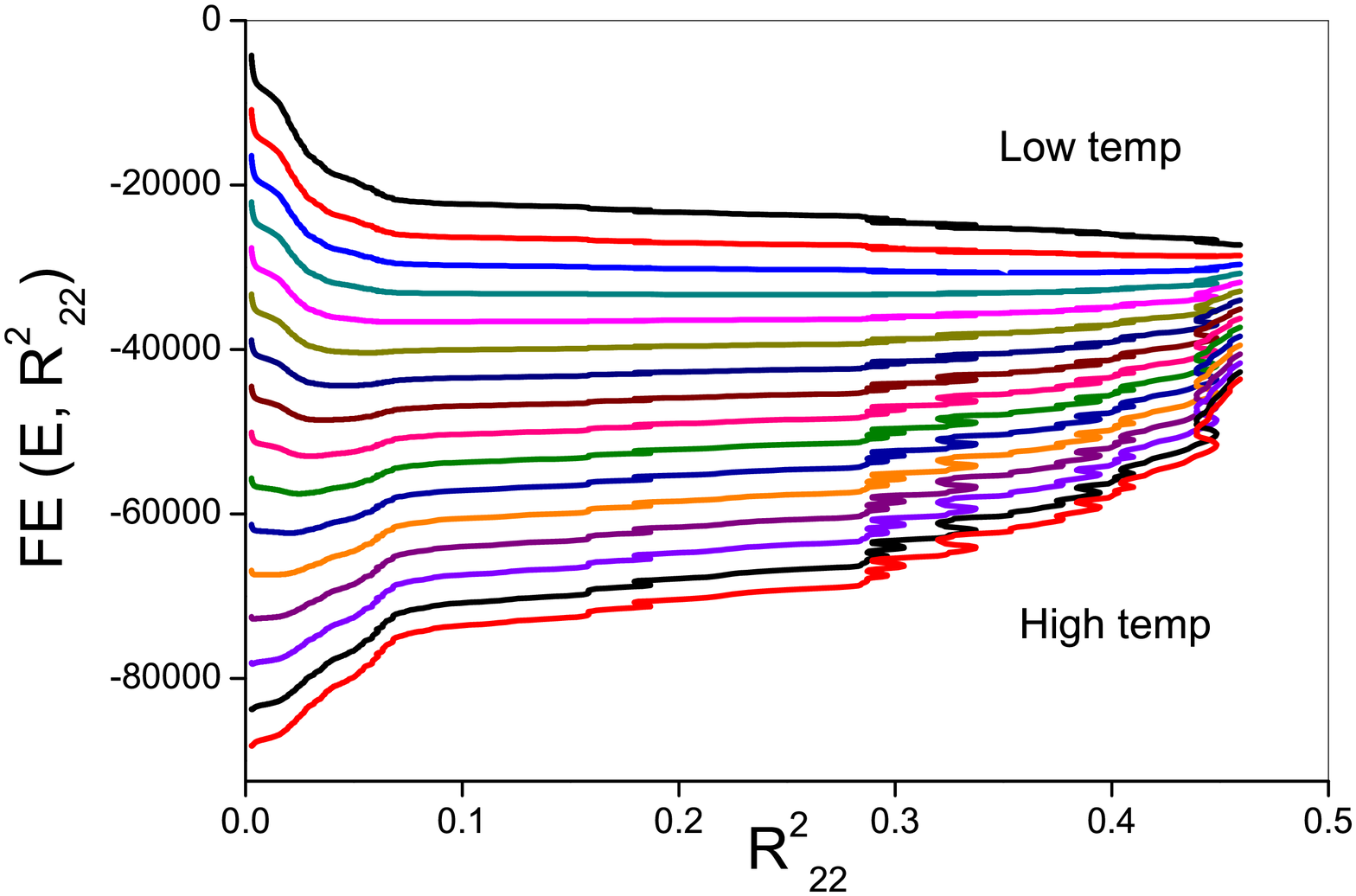}
\label{fig:11c}}
}
\caption{Representative free energy plotted as a function of (a) energy 
(b)$R^{2} _{00}$ (c) $R^{2}_{22}$ at the point  B4 ($\lambda^{'}$ = 
0.674)} 
\label{fig:11}
\end{figure}

\begin{figure}
\centering{
\subfigure[]{\includegraphics[width=0.45\textwidth]{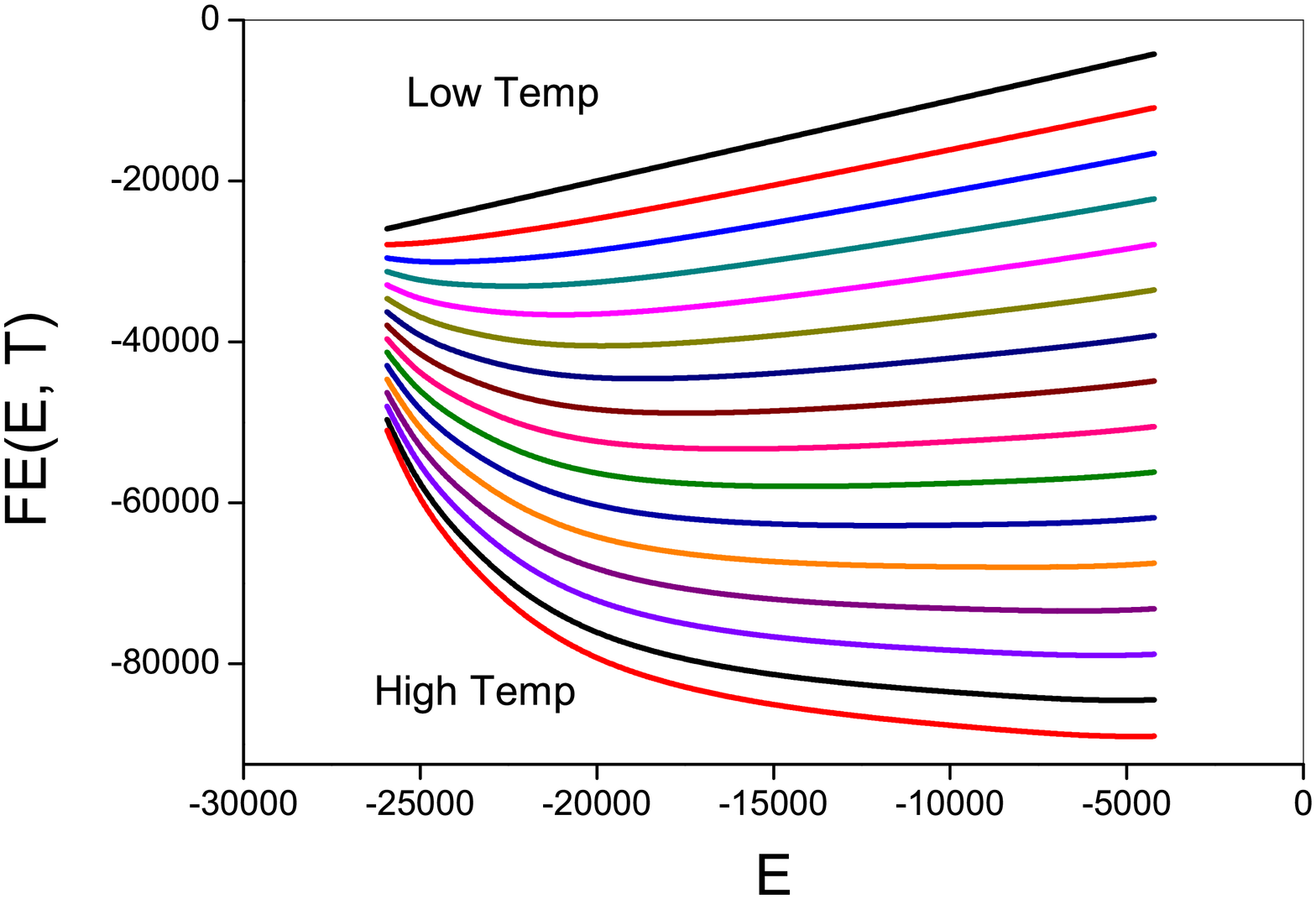}
\label{fig:12a}}
\subfigure[]{\includegraphics[width=0.45\textwidth]{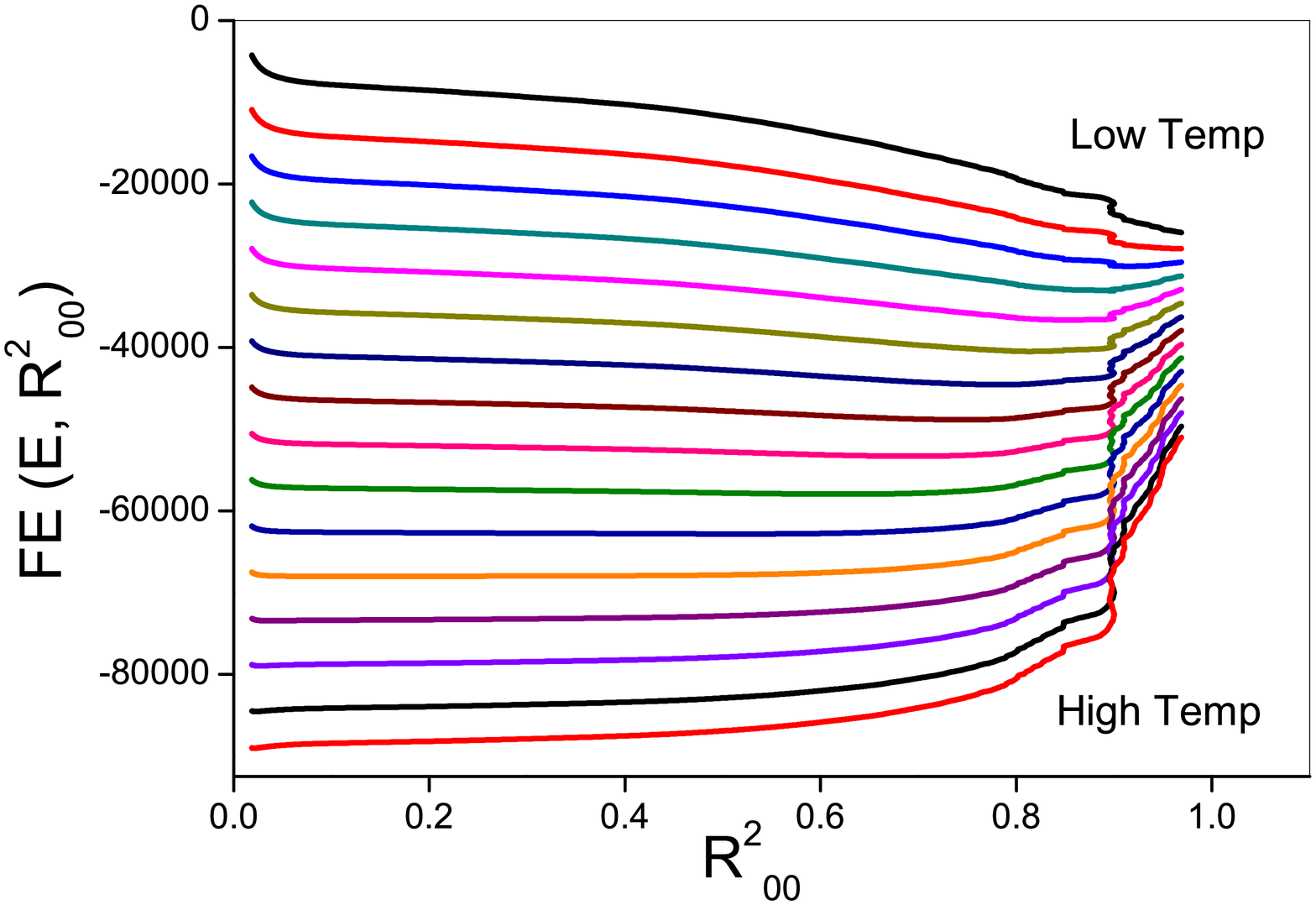}
\label{fig:12b}}
\subfigure[]{\includegraphics[width=0.45\textwidth]{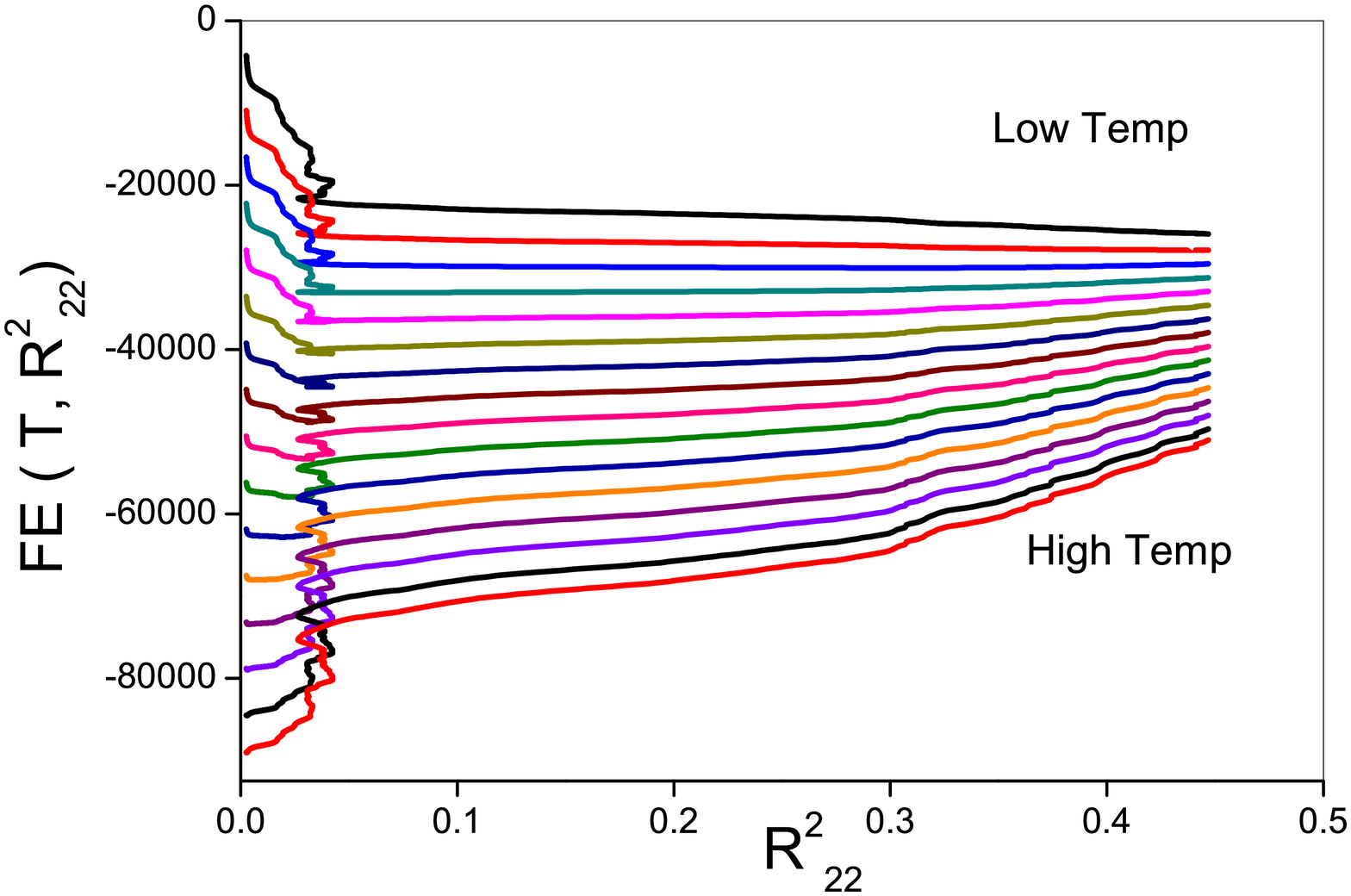}
\label{fig:12c}}
}
\caption{Representative free energy plotted as a function of (a) energy 
(b) $R^{2} _{00}$ (c) $R^{2}_{22}$ at the point Z ($\lambda^{'}$ = 
0.691)}  
\label{fig:12}
\end{figure}

\begin{figure}
\centering{
\subfigure[]{\includegraphics[width=0.45\textwidth]{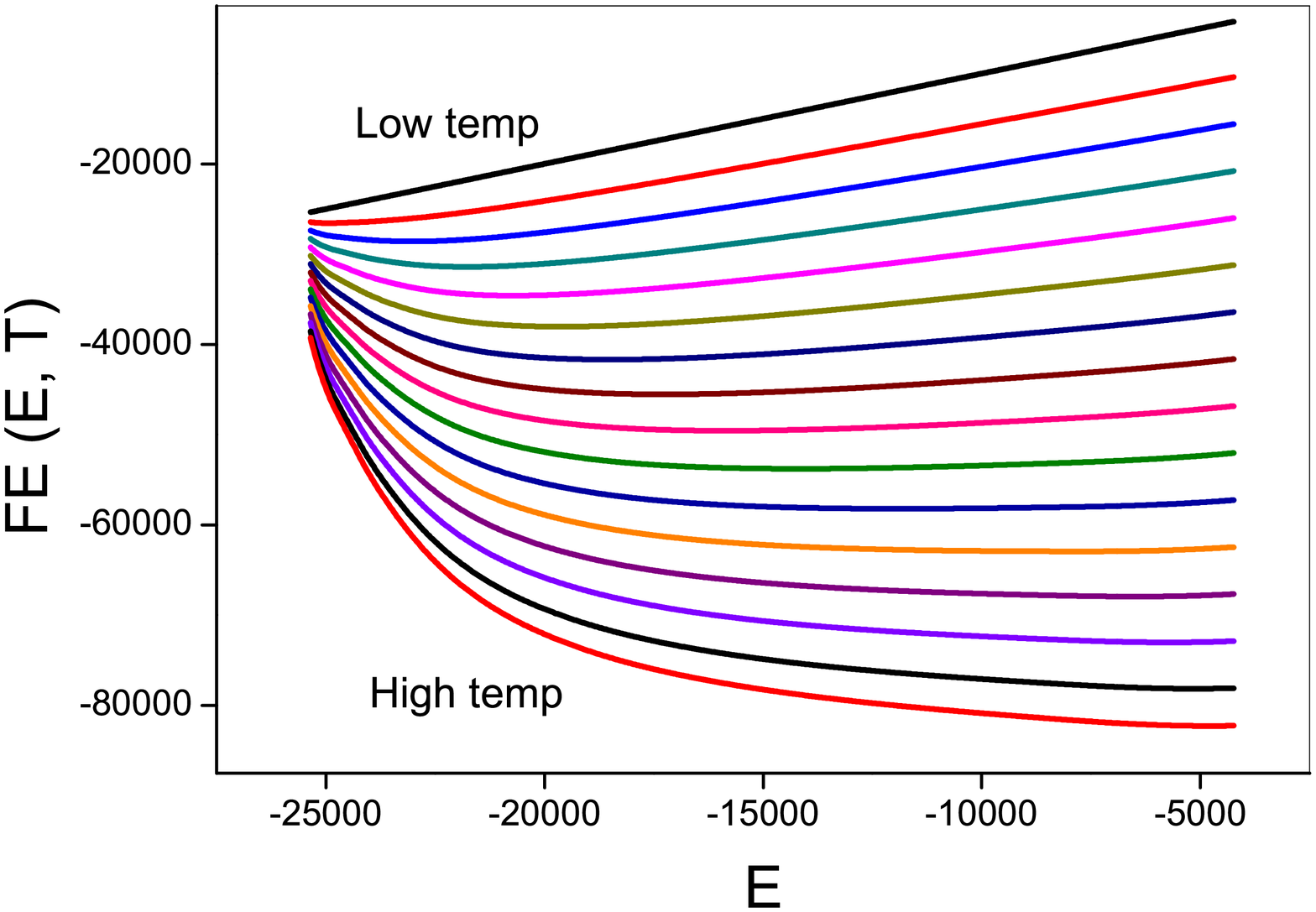}
\label{fig:13a}}
\subfigure[]{\includegraphics[width=0.45\textwidth]{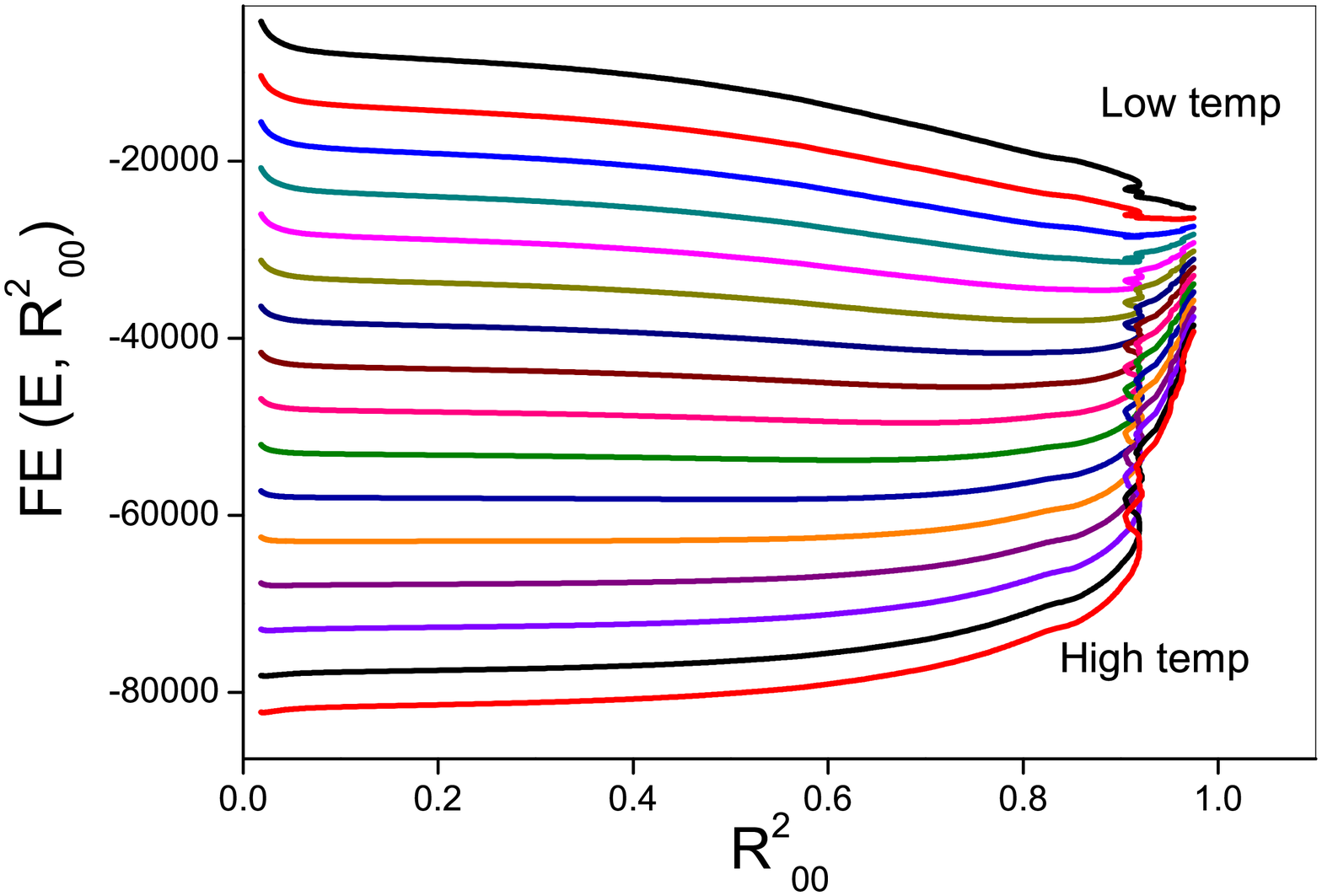}
\label{fig:13b}}
\subfigure[]{\includegraphics[width=0.45\textwidth]{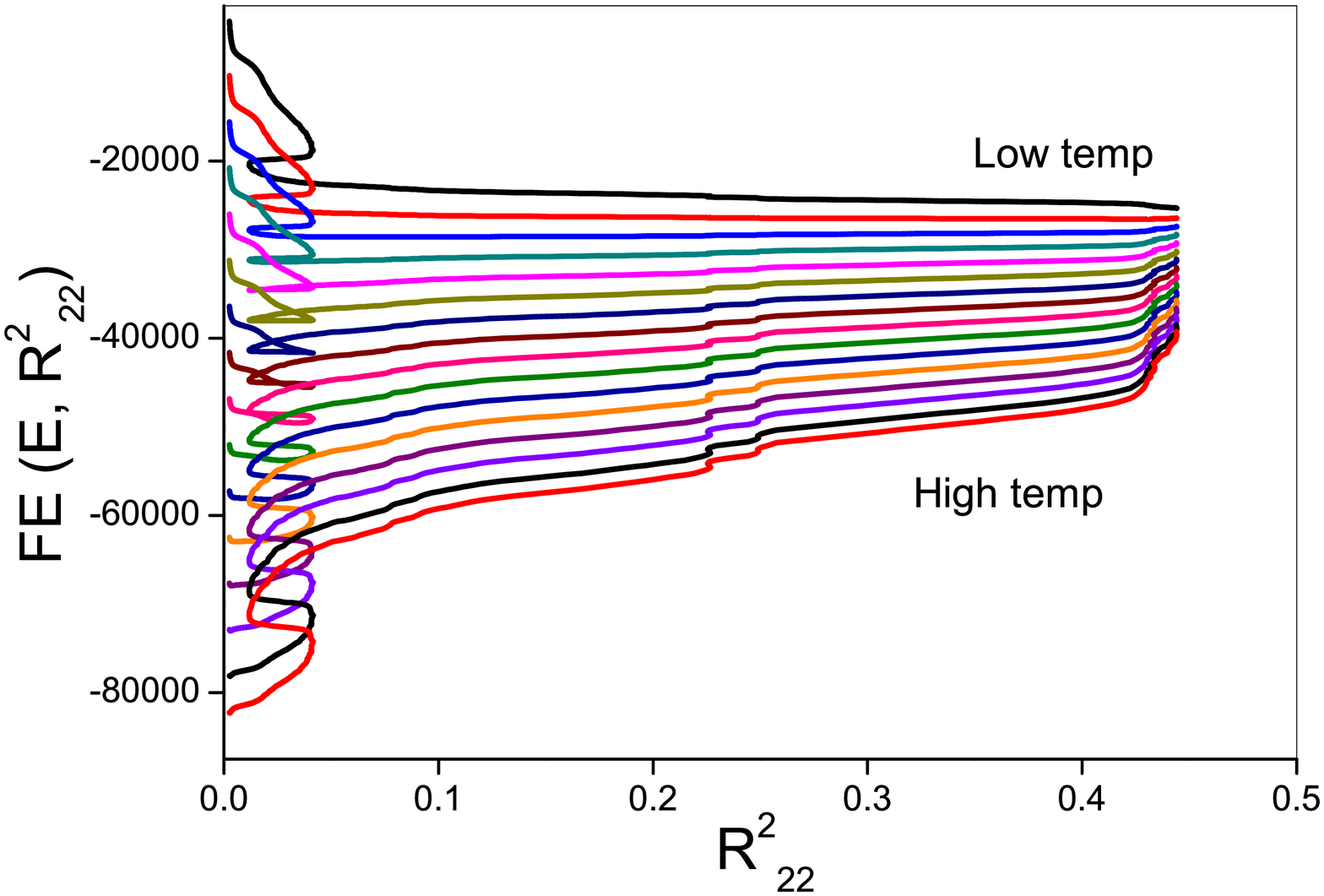}
\label{fig:13c}}
}
\caption{Representative free energy plotted as a function of (a) energy 
(b) $R^{2} _{00}$ (c) $R^{2}_{22}$ at the point B5 ($\lambda^{'}$ = 
0.709)} 
\label{fig:13}
\end{figure}

 The free energy curves obtained for  $\lambda^{'}$= 0.610 ($B_{3}$in the attractive 
 region) are shown in Fig.~\ref{fig:10}. These curves depict the smooth 
 variation of free energy as a function of energy and uniaxial order 
 parameter. However, its variation with respect to $R^{2}_{22}$ shows a 
 small sharp well, (the edge being located at $R^{2}_{22}\simeq 0.02$),
 and the family of curves in Fig.\ref{fig:10c}, as a function of 
 temperature, shows that
 it required significant variation of temperature before the system could shift 
 its free energy minimum away from this restricted region. It appears that 
 during this temperature range, the system accesses microstates with rather 
 small but nonzero degree of biaxiality, constrained however by free energy
 barriers to attain higher degree of biaxiality for considerable range of 
 temperature. This circumstance seems to be manifesting as a corresponding 
 curious variation of $R^{2}_{22}$ at $\lambda^{'}$= 0.610, as in Fig.\ref{fig:}.
 We find that the development of such free energy barriers (with respect
 to $R^{2}_{22}$ at low values) and the requirement of the system to cool 
 sufficiently to overcome them before accessing higher macroscopically
 observable values, is generic. All the data collected in this region 
  supports and corroborates the simulated order parameter
 profiles reported in the previous figures. It is very interesting that 
 such barriers are exhibited only along the path of biaxial order, but not
 along energy or uniaxial order. This implies a complex free energy 
 surface in the 2-d space of order parameter, offering initial barriers to
 a significant development of biaxial order, until the system is 
 sufficiently cooled. Figs. \ref{fig:11} - \ref{fig:13} demonstrate
 this view point.

\subsection{Segment ZW: Range of  $\lambda^{'} $ = (0.691 - 0.747) }

We now present data obtained beyond the point Z (Fig.\ref{fig:e2}).
 In this region, the biaxial-biaxial tensorial 
coupling term $\lambda \rightarrow 0$ asymptotically, leading to a
special case of the interaction Hamiltonian. The case for $\lambda$ = 0 
was studied earlier through simulations \cite{luck80}. It was found that 
in the absence of the biaxial-biaxial interaction term, only a uniaxial 
phase could be obtained on condensation from the isotropic phase. We 
present here the simulation results in the case of $\lambda \rightarrow 0$.
The mean field analysis predicts that the Hamiltonian is partly repulsive
 in this region and excluded volume effects play a major role \cite{Bisi07}.
\begin{figure}
\centering{
\subfigure[]{\includegraphics[width=0.45\textwidth]{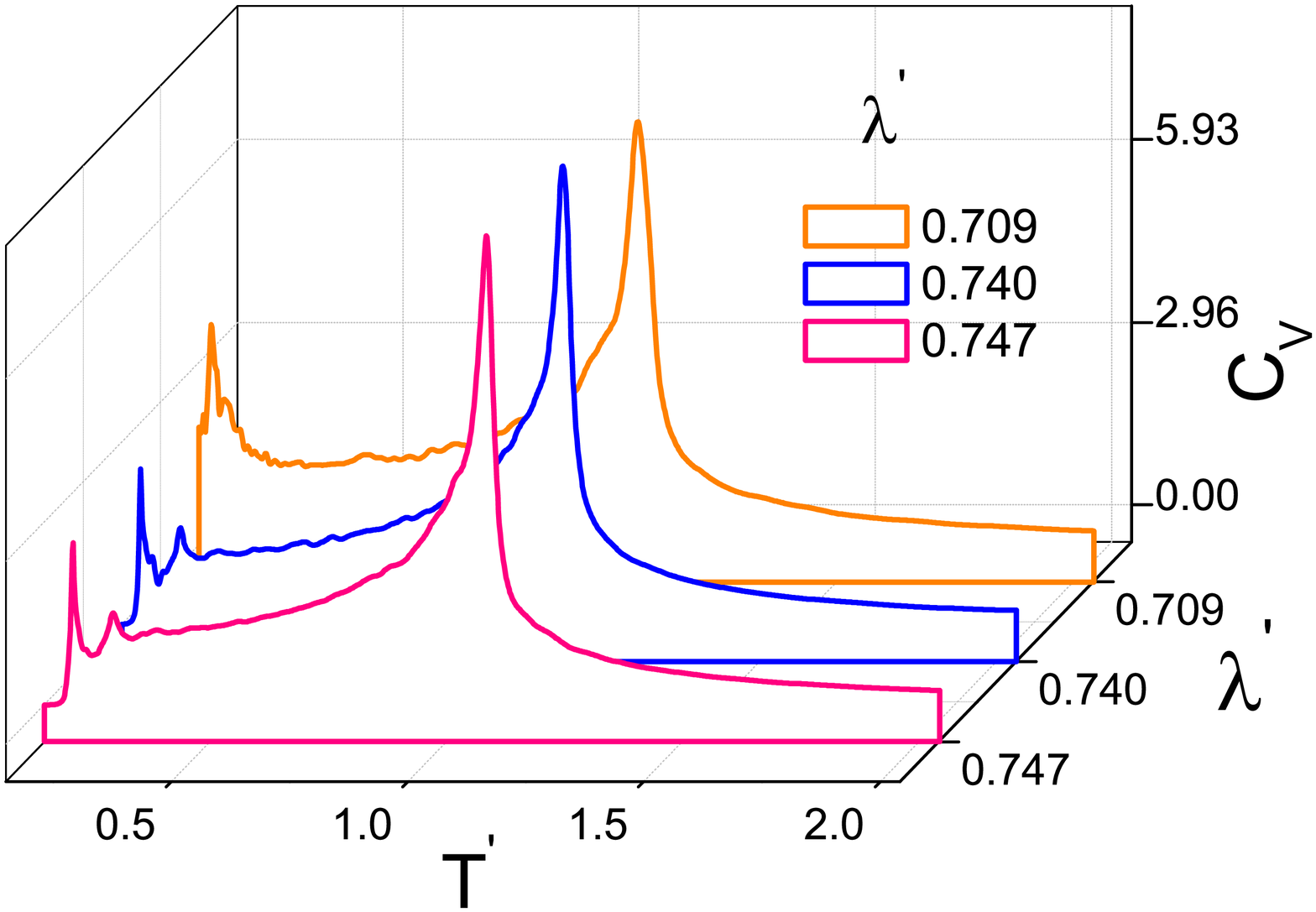}
\label{fig:14a}}
\subfigure[]{\includegraphics[width=0.45\textwidth]{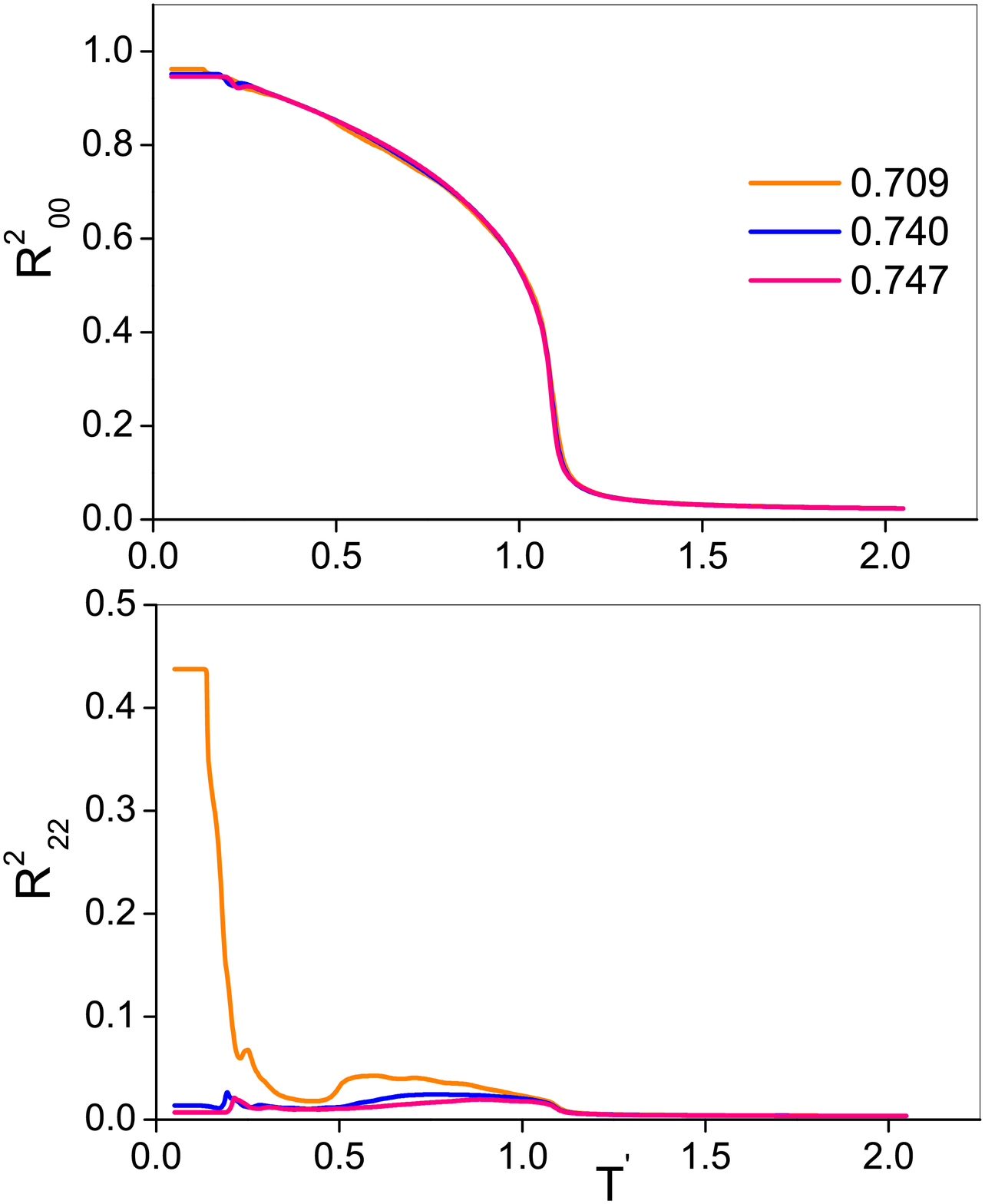}
\label{fig:14b}}
}
\caption{Comparison of (a) specific heat (b) order parameter profiles 
for values of $\lambda^{'}$ in the range 0.709 - 0.747 (L=15) } 
\label{fig:14}
\end{figure}
\begin{figure}
\centering
\includegraphics[scale=0.4]{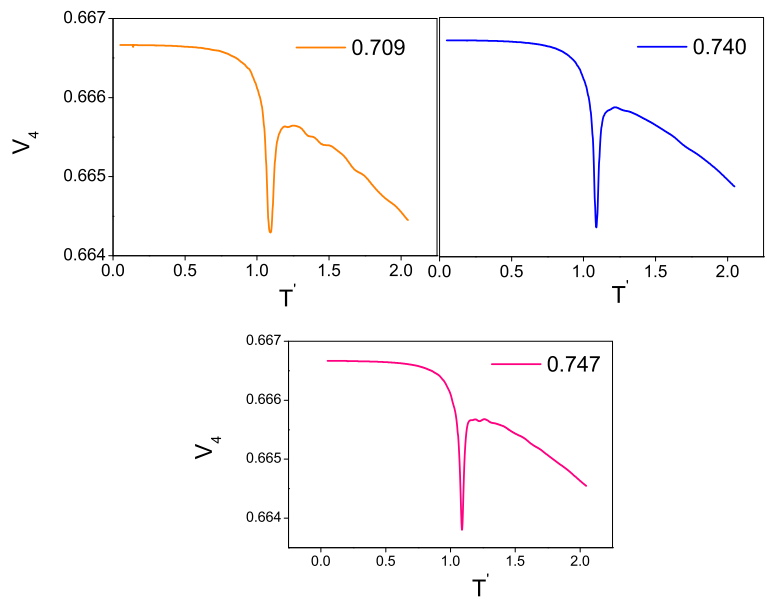}
\caption{Energy cumulant $V_{4}$ for values of  $\lambda^{'}$ in 
the range 0.709 - 0.747 (L=15)} 
\label{fig:15}
\end{figure} 
Due to the constraints imposed by computational time, we could obtain 
data in this range of $\lambda^{'}$ only for a smaller system, with L=15.
(instead of L=20, as in the earlier case). The specific heat and order 
parameter profiles are depicted in Figs.~\ref{fig:14a} and  \ref{fig:14b}. 
The energy cumulants $V_{4}$ are shown in Fig.~\ref{fig:15}. It may be 
observed that  the specific heat profiles show evidences of two transitions.
 The order parameter profiles depict the onset and 
growth of uniaxial order at $T_{1}$  for all values of $\lambda^{'}$. 
The biaxial order parameter increases at $T_{2}$ (in the biaxial phase)
 for $\lambda^{'}$ = 0.709, but remains close to zero for $\lambda^{'}$
 = 0.740 and 0.747. This behaviour is as expected from mean field 
 considerations at such values of $\lambda^{'}$, very close to the base OW.
\begin{figure}
\centering{
\subfigure[]{\includegraphics[width=0.45\textwidth]{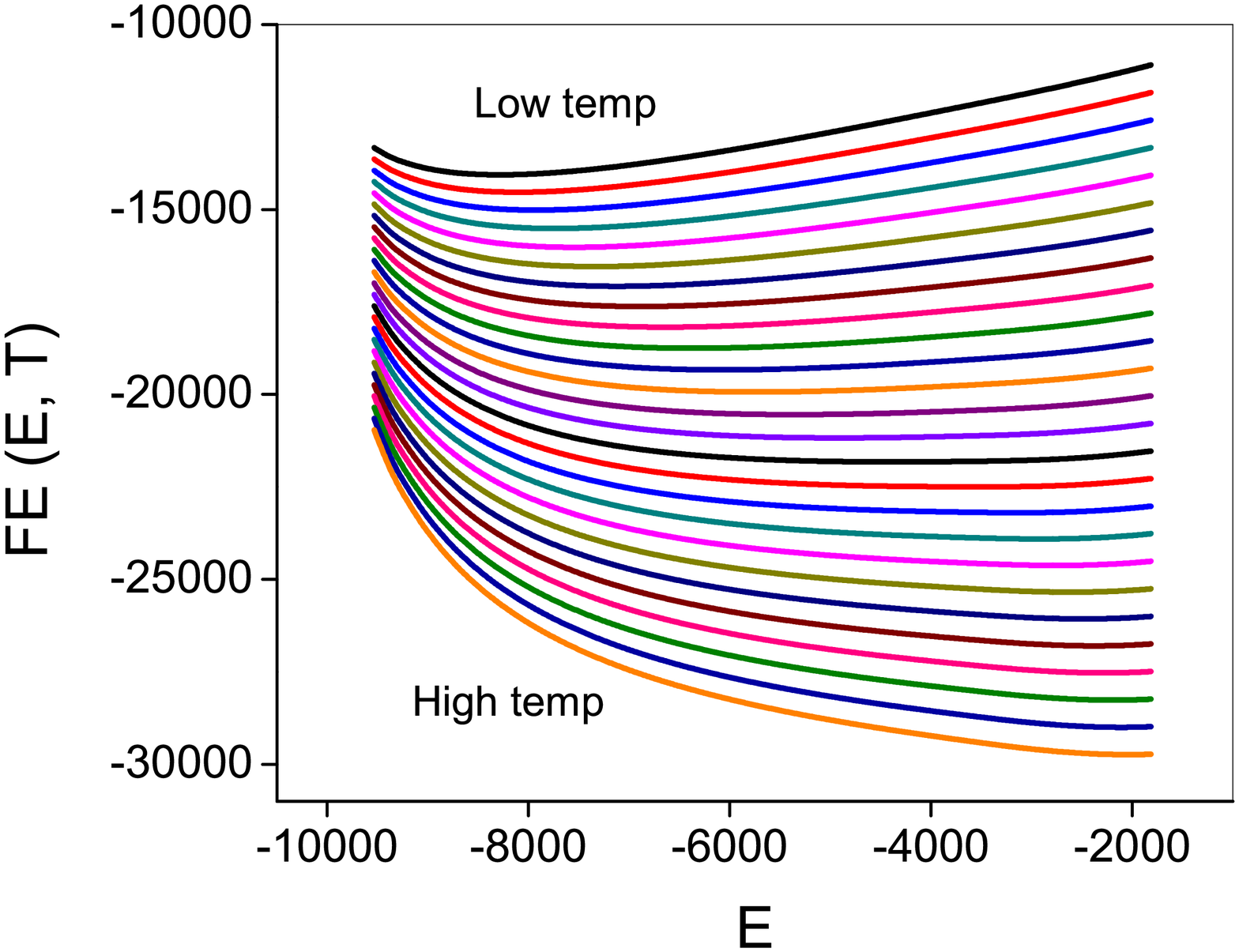}
\label{fig:16a}}
\subfigure[]{\includegraphics[width=0.45\textwidth]{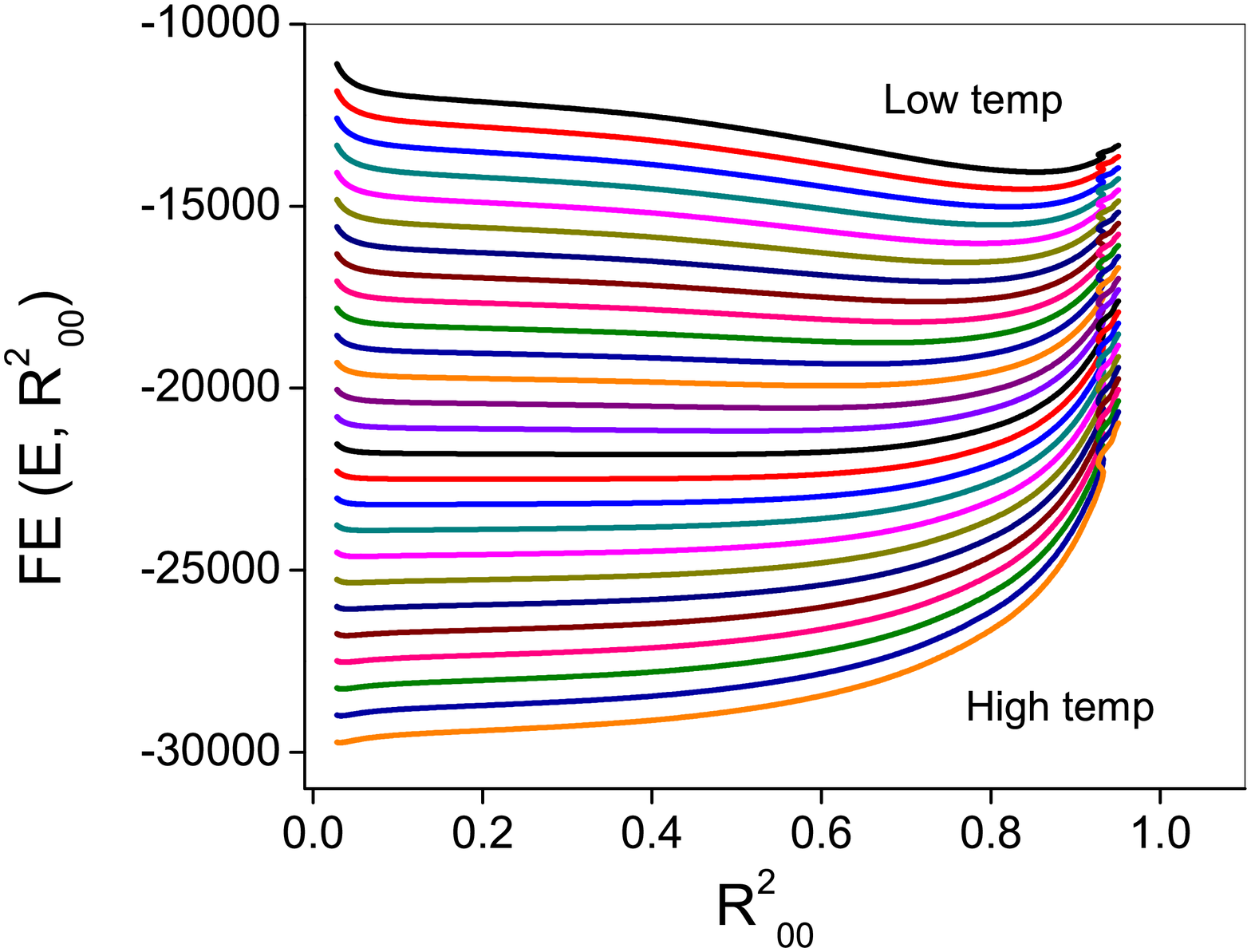}
\label{fig:16b}}
\subfigure[]{\includegraphics[width=0.45\textwidth]{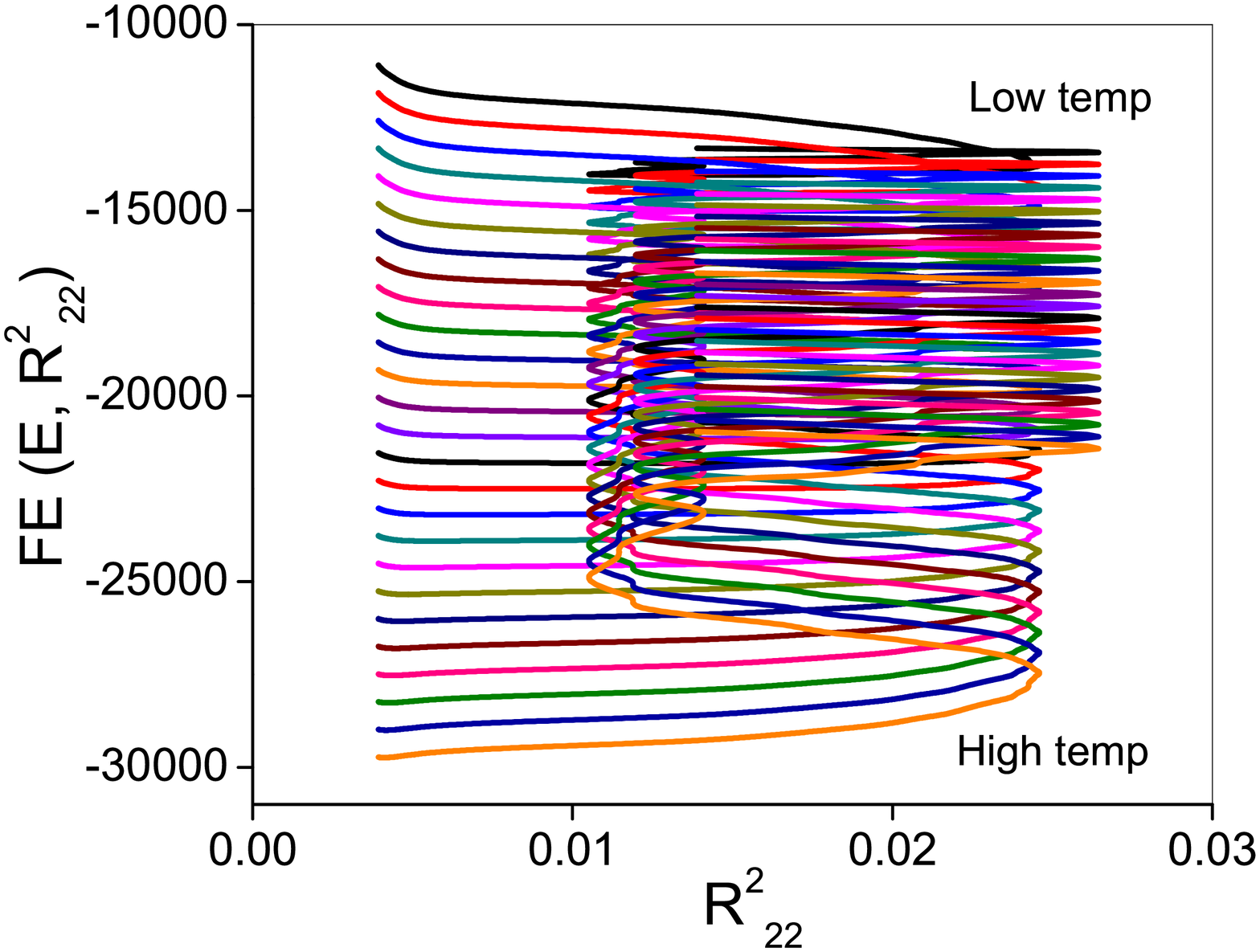}
\label{fig:16c}}
}
\caption{ Representative free energy plotted as a function of (a) energy 
(b)$R^{2} _{00}$ (c) $R^{2}_{22}$ at point B6 ($\lambda^{'}$ = 0.740) 
} 
\label{fig:16}
\end{figure}
The free energy plots for $\lambda^{•}$ = 0.740
are shown in Figs.~\ref{fig:16a} - \ref{fig:16c}. The free energy variation 
with respect to  $R^{2}_{22}$ for $\lambda^{'}$ = 0.740 again confirms the 
presence of barriers for the growth of biaxial order at points close to
 the base OW. The biaxial state is obviously not stable at such parameter
 points of the Hamiltonian.

\section{Conclusions}
In this paper,  the interior of essential triangle is explored 
with entropic sampling method along a trajectory IW, a 
 line drawn from apex I of the triangle to the mid point W of the base OV
(Fig. \ref{fig:1}). We refer to the arc length OIW as $\lambda^{'}$, 
for purposes of discussion. As per MF treatment this line cuts both the trajectories
of $C_{2}C_{3}$ and $C_{1}C_{3}$. Thus the phase sequences along this line IW 
should be qualitatively similar to that on the $\lambda$-axis. In particular the
direct transition from isotropic to biaxial phase is expected to be interrupted by
a uniaxial nematic phase beyond a value as $\gamma$ increases, and the temperature
range of the uniaxial phase should progressively increase, suppressing the second
transition temperature, till the line cuts the parabola, at Z (Fig.\ref{fig:1}).
Results from mean field treatment in the partly repulsive region on this 
line IW are not available for direct comparison, even though it is established 
that biaxial phase would not be stable  at the point W \cite{matteis07, matteis08}.
The phase sequences in the present work qualitatively follow this scenario, but with 
a curious deviation. The intervening "uniaxial " phase $N_{U}$ is not 
strictly devoid of biaxial symmetry. Indeed all along the line, beyond 
$K^{'}$ (Fig.\ref{fig:e1}), and upto point
Z, the onset of the uniaxial order is invariably accompanied by a small, but
unmistakable, development of biaxial symmetry. We thus refer to this 
phase as $N_{U^{'}}$, to make this subtle distinction. This small degree
 of biaxiality of the $N_{U^{'}}$ phase is temperature independent within that
 phase, and is also fairly independent of its location in the trajectory 
 beyond $K^{'}$.
 An examination of the  free energy profiles, drawn as a function of both
 the major order parameters, show interesting features: while the free energy
 curves show smooth variation of the minima with respect to $R^{2}_{00}$ as the 
 temperature is varied, the case of $R^{2}_{22}$ is qualitatively different. 
 These profiles exhibit free energy barriers at low values of $R^{2}_{22}$, 
 which could be overcome (thereby pushing the system to access regions of
 higher and discernible order), only after these initial barriers could be 
 overcome on considerable cooling. Thus these results 
 show a complex free energy surface that develops with decrease of temperature
 on a typical trajectory inside the triangle. It appears that development of
 a $N_{U^{'}}$ phase with a small biaxial order ($\leq 0.05)$ is expected, 
 and the degree of this symmetry is restricted by the free energy barriers 
 till the system is permitted to access these regions of biaxial order. 
 Given that such barriers are strongly dependent on the size of the system, 
 it is a plausible conjecture to suggest that in real systems these barriers 
 are not readily overcome (or equivalently, requires significant cooling of 
 the medium), and hence their biaxial order appears to be restricted inherently.
 Under such circumstances requiring wider temperature ranges to overcome 
 barriers, real systems may have other competing interactions (like translational
 degrees, influencing the phase sequence qualitatively differently , e.g
 layer formation). Deviations of real systems from MF predictions \cite{Bisi08} 
 could perhaps be understood in these terms.


\begin{thebibliography}{}
\bibitem{freiser} M. J. Freiser, Phys. Rev. Lett. \textbf{24}, 1041 (1970).

\bibitem{Straley} J. P. Straley, Phys. Rev. A  \textbf{10}, 1881 (1974).

\bibitem{Remler} D. K. Remler, and A. D. J. Haymet, J. Phys. Chem., 
\textbf{90}, 5426 (1986).

\bibitem{Luck75} G. R. Luckhurst, C. Zannoni, P. L. Nordio, and U. Segre, 
Mol. Phys. \textbf{30}, 1345 (1975).

\bibitem{Sonnet}A. M. Sonnet, E. G. Virga, and  G. E. Durand, 
Phys. Rev. E \textbf{67}, 061701 ( 2003).

\bibitem{matteis05A}G. De Matteis, and E. G. Virga, Phys. Rev. E \textbf{71},
 061703 (2005).
 
\bibitem{Bisi06} F. Bisi, E. G. Virga, E. C. Gartland Jr., G. De Matteis,
 A. M. Sonnet, and G. E. Durand, Phys. Rev. E \textbf{73}, 
 051709 (2006).
 
 \bibitem{Bisi07} F. Bisi, S. Romano, and E. G. Virga,  Phys. Rev. E \textbf{75},
  041705 ( 2007)
 
\bibitem{matteis07} G. De Matteis, F. Bisi, and E. G. Virga,
Continuum. Mech. Thermodynamics. \textbf{19}, 1 (2007).


  \bibitem{Alben} R. Alben, Phys. Rev. Lett. \textbf{30}, 778 (1973).       
 
\bibitem{Bocara} N. Bocara, R. Mejdani, and L. De Seze, 
J. Phys. (Paris) \textbf{38}, 149 (1976).

\bibitem{Gramsbergen} E. F. Gramsbergen, L. Longa, and  W. H. de Jeu,
Phys. Rep. \textbf{135}, 195 (1986).

\bibitem{Allender08} D. Allender and L. Longa, Phy. Rev. E 
\textbf{78}, 011704 (2008).

\bibitem{Mukherjee09} P. K. Mukherjee and Kallol Sen, J. Chem. Phys. 
\textbf{130}, 141101 (2009).

\bibitem{Yu}L. J. Yu and A. Saupe, Phys. Rev. Lett. \textbf{45}, 1000 (1980).
 
\bibitem{Acharya04} B. R. Acharya, A. Primak and S. Kumar, 
Phys. Rev. Lett.  \textbf{92}, 145506 (2004).

\bibitem{madsen} L. A. Madsen, T. J. Dingemans, M. Nakata, and  E. T. Samulski, 
Phys.  Rev. Lett. \textbf{92}, 145505  (2004).
 
\bibitem{Merkel} K. Merkel, A. Kocot, J. K. Vij, R. Korlacki, G. H. Mehl, 
and T. Meyer, Phys. Rev. Lett.  \textbf{93}, 237801 (2004).

\bibitem{Figueirinhas05} J. L. Figueirinhas, C. Cruz, D. Filip, G. Feio, 
A. C. Ribeiro, Y. Frere, T. Meyer, and G. H. Mehl, Phys. Rev. Lett.
 \textbf{94}, 107802  (2005). 

\bibitem{Severing} K. Severing and K. Saalwachter, Phys. Rev. Lett.  \textbf{92},
 125501  (2004).
 
\bibitem{luck80} G. R. Luckhurst and  S. Romano, Mol. Phys. 
\textbf{40}, 129 (1980).

\bibitem{Allen90} M. P. Allen. Liq. Cryst. \textbf{8}, 499 (1990).

\bibitem{Biscarini95} F. Biscarini, C. Chiccoli, P. Pasini, F. Semeria,
 and C. Zannoni, Phys. Rev. Lett. \textbf{75}, 1803 (1995).
 
\bibitem{chiccoli99A} C. Chiccoli, P. Pasini, F. Semeria, and  C. Zannoni,
 Int. J. Mod. Phys. C \textbf{10}, 469 (1999).

\bibitem{Berardi03} R. Berardi and C. Zannoni,
Mol. Cryst. Liq. Cryst. \textbf{396}, 177 (2003).

\bibitem{BerardiB} R. Berardi, L. Muccioli, S. Orlandi, M. Ricci and
 C. Zannoni, J. Phys.: Condens. Matter \textbf{20}, 463101 (2008).
 
\bibitem{matteis05B}G. De Matteis, S. Romano, and E. G. Virga, 
Phys. Rev. E \textbf{72}, 041706 (2005).

\bibitem{kamala14} B. Kamala Latha, Regina Jose, K. P. N. Murthy and V. S. S. Sastry, Phys. Rev. E \textbf{89}, 050501(R) (2014).

\bibitem{kamala15}B. Kamala Latha, Regina Jose, K. P. N. Murthy and V. S. S. Sastry, Phys. Rev. E \textbf{92}, 012505 (2015).

\bibitem{Wang}F. Wang and D. P. Landau, Phys. Rev. Lett.
\textbf{86}, 2050 (2001); F. Wang and D. P. Landau, Phys. Rev. E
\textbf{64}, 056101 (2001).

\bibitem{Zhou} C. Zhou, T. C. Schulthess, S. Torbrugge, and D. P. Landau, 
Phys. Rev. Lett. \textbf{96}, 120201 (2006).

\bibitem{Jayasri09}D. Jayasri, Ph. D Thesis, \textit{Non-Boltzmann Monte Carlo study of Confined Liquid Crystals and Liquid Crystal Elastomers}, University of Hyderabad, India (2009).


\bibitem{romano} S. Romano, Physica A \textbf{337}, 505 (2004).

\bibitem{Landau1}D. P. Landau and K. Binder, \textit{A Guide to Monte Carlo Simulations in Statistical Physics}, Cambridge University Press, 2nd edition(2005).

\bibitem{Murthy1}K. P. N. Murthy, \textit{Monte Carlo Methods in Statistical Physics} Universities Press, India(2004).


\bibitem{Rathore03}N. Rathore, T. A. Knotts and J. J. de Pablo, Biophysics. J. \textbf{85}, 3963 (2003).


\bibitem{Seaton10}D. T. Seaton, T. Wust and D. P. Landau, \textit{Phys. Rev. E} \textbf{81}, 011802 (2010).

\bibitem{Priya11} Priya Singh, Subir. K. Sarkar and Pradipta Bandyopadhyay,
Chem. Phys. Lett. \textbf{514}, 357 (2011).


\bibitem{Poulain}P. Poulain, F. Calvo, R. Antoine, M. Broyer and P. Dugourd, 
Phys. Rev. E \textbf{73}, 056704 (2006).

\bibitem{Sinha09}S. Sinha and  S. K. Roy, Phys. Lett. A \textbf{373}, 308 (2009).

\bibitem{Raj} Raj Shekhar, Jonathan K. Whitmer, Rohit Malshe, J. A. Moreno-Razo,
Tyler F. Roberts and Juan J. de Pablo, J. Chem. Phys. \textbf{136}, 234503(2012). 

\bibitem{Yang13} Yang Wei Koh and Hwee Kuan Lee, Phys. Rev. E \textbf{88}, 053302 (2013).

\bibitem{Vogel13} T. Vogel, Y. W. Li, T. Wust and D. P. Landau,
Phys. Rev. Lett. \textbf{110}, 210603 (2013).

\bibitem{Katie14} Katie A. Maerzke, Lili Gai, Peter T. Cummings and Clare McCabe,
J. Phys. Conference Series \textbf{487}, 012002 (2014).

\bibitem{Xie14} Y. L. Xie, P. Chu, Y. L. Wang, J. P. Chen, Z. B. Yan, J. -M. Liu,
Phys. Rev. E \textbf{89}, 013311 (2014).

\bibitem{Landau13} Lili Gai, Thomas Vogel, Katie A. Maerzke, Christopher 
R. lacovella, David P. Landau, Peter T. Cummings and Clare McCabe,
J. Chem. Phys. \textbf{139}, 054505 (2013).


\bibitem{jayasri}D. Jayasri, V. S. S. Sastry, and K. P. N. Murthy,
Phys. Rev. E \textbf{72}, 036702  (2005).

\bibitem{LL}P. A. Lebwohl and  G. Lasher, Phys. Rev. A \textbf{6}, 426 ( 1972).


\bibitem{Swensden} R. H. Swendsen and J. S. Wang, Phys. Rev. Lett. \textbf{58}, 86 (1987).


\bibitem{Binder} K. Binder, Z. Physik. B \textbf{43}, 119 (1981); 
K. Binder, Phys. Rev. Lett. \textbf{47}, 693 (1981).


\bibitem{Robert} Robert J Low, Eur. J. Phys. \textbf{23}, 111 ( 2002).

\bibitem{matteis08} G. De Matteis and S. Romano, Phys. Rev. E \textbf{78},
021702 (2008).

\bibitem{Bisi08}F. Bisi, G. R. Luckhurst, E. G. Virga, \textit{Phy. Rev. E} \textbf{78}, 021710 (2008).

\end{thebibliography}
\end{document}